\documentclass[%
 reprint,
 amsmath,amssymb,
 aps,
 prapplied
]{revtex4-2}

\usepackage{comment}
\usepackage{xcolor}
\usepackage{amsmath}
\usepackage{graphicx}
\usepackage{dcolumn}
\usepackage{bm}

\usepackage{fancyhdr}

\begin{document}

\preprint{APS/123-QED}

\title{
Steady-State Exceptional Point Degeneracy and Sensitivity of Nonlinear Saturable Coupled Oscillators 
}

\author{Benjamin Bradshaw}
\affiliation{Department of Electrical Engineering and Computer Science, University of California, Irvine, California 92697, USA}

\author{Amin Hakimi}
\affiliation{Department of Electrical Engineering and Computer Science, University of California, Irvine, California 92697, USA}

\author{Filippo Capolino}
\email{f.capolino@uci.edu}
\affiliation{Department of Electrical Engineering and Computer Science, University of California, Irvine, California 92697, USA}

\begin{abstract}

A coupled oscillator system displays enhanced sensitivity of its saturated steady-state (SS) oscillation frequency to small parameter perturbations near an exceptional point degeneracy (EPD), a property that can be used to realize EPD-based sensors. Linear $\mathcal{PT}$-symmetric systems, consisting of two coupled resonators, exhibit EPDs around which square-root sensitivity is observed. However, linear models are insufficient for realistic systems that rely on nonlinear, saturable gain elements, particularly when $\mathcal{PT}$-symmetry is broken. Thus, we study the SS of a general system of two coupled oscillators featuring EPDs and saturable nonlinear gain, using coupled-mode theory. We do this by synthesizing and extending prior SS analyses of the system's stability, and its square-root and cubic-root oscillation frequency sensitivity at a unique third-order SS-EPD. We include an SS analysis of the saturated gain values, energy, and the oscillation frequency's sensitivity in the vicinity of the third-order SS-EPD, providing a comprehensive analysis of the system's various SS regimes. We determine that the stable and bistable regions in parameter space directly depend on the saturated gain values; that the dynamic range of high sensitivity around degenerate conditions is extended  by increasing losses, consequently reducing the system's stored energy; and that, to exploit the cubic-root-like sensitivity associated to the third-order SS-EPD, the suggested working regime is best confined to operation within the weakly coupled regime and not exactly at the third order SS-EPD. Finally, we apply the model to two electronic circuits that exhibit cubic-root sensitivity, demonstrating the application and limitations of this analysis.
\end{abstract}

\maketitle

\section{Introduction}

\begin{figure}[t]
\begin{centering}
\includegraphics[width=2in]{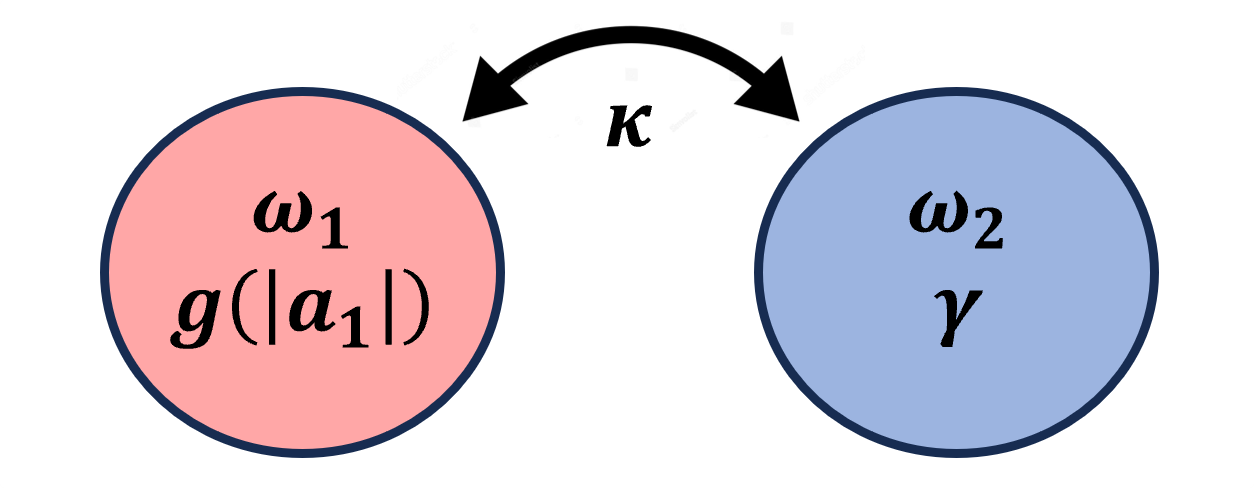}
\par\end{centering}
\caption{Generic coupled-oscillator system made of two resonators resonating at $\omega_1$ and $\omega_2$,  with saturable nonlinear gain $g$ and loss $\gamma$, respectively, and coupling $\kappa$. \label{fig:CoupledOsc}}
\end{figure}

Oscillating systems are a fundamental building block of many modern devices. Coupled oscillators and their behavior have been studied extensively for several decades \cite{Alder1946_CoupOsc}, and research on this fundamental topic remains prevalent today due to their complex behavior caused by nonlinearity. Coupled oscillators have applications in current research fields such as quantum computing \cite{Hayato2019_KerrQComputing,Chmielewski2025_ThreeCoupledKerr}, wireless power transfer, wireless sensing, and others. Individual and coupled oscillators exhibit complex behaviors, ranging from simple harmonic motion to the complex and chaotic dynamics found in nonlinear systems such as the Van der Pol oscillator \cite{Pol1934_VanDerPolOsc}.
To obtain stable oscillations in electronics, negative resistive elements or positive feedback are added to balance natural losses in a circuit \cite{Razavi1998_RFElectronics}. Gain elements are used in microcavities \cite{Zhou2016_PTSymmBreak} and lasers \cite{Siegman1986_Lasers}, and are generally described as nonlinear active gain elements. The steady-state (SS) regime  for a single oscillator with an active gain element is trivial, as it occurs when the gain saturates to a value that balances the inherent losses of the system. However, when an oscillator with a nonlinear active gain component is coupled to another oscillator, the behavior becomes complex. In this paper, we investigate a basic system of two coupled oscillators, one of which contains a nonlinear active gain element, as depicted in Fig.~\ref{fig:CoupledOsc}, and focus on the sensitivity of the SS oscillation frequency to system perturbations.

In parallel, over the past decades, the topic of exceptional point degeneracies (EPDs) has gained significant interest because of their unique physics, as in  \cite{Kato1966_perturbation, Heiss1990Avoided,Sun1990MultipleEigenvSens,seyranian1993sensitivity,Heiss2000Repulsion, dembowski2001experimental, heiss2004exceptionalP,Seyranian_2005Coupling}, and more recently due to the growing topic of parity-time ($\mathcal{PT}$)-symmetric physics \cite{Bender2002Generalized,Ruter2010Observation,Zheng2010_OptPTSym,Schindler2011_LRCPTSymm,Heiss2012Thephysics, Hodaei2014Parity, Ruter2010Observation,Peng2014_MicroPTSymm,chang2014_parity,Milburn2015_linearEPD,Ge2016_nonlinear,miri2019exceptional, krasnok2021parityTSymm}. 
In most of these investigations, EPDs are observed in a linear system made of two coupled resonators (as shown in Fig.~\ref{fig:CoupledOsc}, though here we focus on nonlinear dynamics), where eigenvalues and eigenvectors of the linear system coalesce. The letter “D” in EPD emphasizes the key physical concept of “degeneracy” of two eigenmodes of a linear system \cite{Berry2004Physics}, described in terms of eigenvalues and eigenvectors. 
For several years, EPDs have been well studied for their unique characteristics associated with the square-root-like enhanced sensitivity to system's perturbations \cite{wiersig2014enhancing,chen2017exceptional,Mortensen2018_FluctuationsPTSymm,chen2018generalized,chen2019sensitivityNearEP,  wiersig2020prospects, wiersig2020review,rosa2021exceptional, Binkowski2024_WiersigEPSensitivity}, a property related to the Puiseux fractional power expansion of an eigenvalue (or eigenvector) for an EPD-system's small perturbation \cite{welters2011explicit}. It has been shown that EPDs can be achieved in other topologies as well, such as two resonators with negative inductance and capacitance \cite{rouhi2022high,Rouhi2022Exceptional,rouhi2024simpleReciprCirc, nikzamir2025exceptionalGyrat}, using a single resonator with a time-varying component \cite{Kazemi2019_LinearPTEPD,kazemi2020ultra,kazemi2022experimental,nikzamir2023timeModulVibra}, as summarized in \cite{Nikzamir2022Achieve}, and also by using nonreciprocal or nonsymmetric coupling between two resonators \cite{wiersig2008asymmetric,peng2016chiral,longhi2017unidirectional,ren2018unidirectional}.
 
The theory of EPD in linear systems clearly shows that there is an increase in sensitivity in the shift of the resonant frequency, compared to resonators without EPD, when a system's parameter is perturbed; this property has been proposed as a general strategy to greatly enhance the sensitivity of sensors. However, it has been debated that working at an EPDs also enhances noise in the system. 
  
A drawback of using {\em linear} systems with EPDs for sensing arises when one of the resonant frequencies, $\omega_2$ or $\omega_1$, of the two resonators is changed (intentionally or unintentionally), because the system loses its $\mathcal{PT}$ symmetry and the two resonances of the coupled resonators become complex valued, with one causing the system's instability, i.e., causing the signal to grow exponentially. Such deviation from ideal $\mathcal{PT}$ symmetry occurs when one of the two resonators' resonances is used for sensing, by perturbing the permittivity of a capacitor or of a ring resonator. 
However, even when the sensing scheme is based on perturbing the coupling between the two resonators, $\mathcal{PT}$ symmetry is achieved only in theory as in practice it is extremely difficult to maintain the exact loss/gain symmetry due to component or fabrication tolerances in electronics or optics.

To overcome the aforementioned instability problems, in \cite{Kazemi2022_CTLPTSymm, Nikzamir2022_HighlySensitive} the authors proposed a solution that relies on exploiting the instability, rather than controlling or mitigating it, which directly inspired this work. They suggested using {\em nonlinear saturable} gain at the steady-state (SS) saturated regime. Indeed, \cite{Nikzamir2022_HighlySensitive} experimentally demonstrated that the SS oscillation frequency is highly sensitive to perturbations of one resonator's resonant frequency (by varying a capacitor), and that the observed oscillation frequency variation exhibits a cubic-root-like sensitivity. Furthermore, the linewidth of the oscillation frequency in \cite{Nikzamir2022_HighlySensitive} was very narrow with low phase noise, allowing an easy measurement of very small frequency shifts. This latter property was experimentally demonstrated also in \cite{nikzamir2025exceptionalGyrat}, where the same nonlinear-saturable gain concept was applied. A theoretical proof of this behavior was not provided in \cite{Nikzamir2022_HighlySensitive}, which is the objective of the current paper. Recently, high sensitivity to perturbations using saturable gain and SS oscillations has also been demonstrated {\em experimentally} in other works, including \cite{Suntharalingam2023_EPVolt,Bai2023_NonlinearityES,Bai2024_Observations,li2024enhancedSens,chen2024inductor,Moncada2025FDanalysisOscillEPD,Moncada2025sensorsOscill}; among these, cubic-root sensitivity was shown in \cite{Bai2023_NonlinearityES, Bai2024_Observations,chen2024inductor, Moncada2025sensorsOscill}.    

Coupled oscillators with nonlinear saturable gain and EPD have also been used in robust wireless power transfer systems, as shown in  \cite{Assawaworrarit2017_robust,assawaworrarit2020robustWPT, Kananian2020Coupling, kananian2022robust, Mohseni2025_WPT,Dong2021_BistableWPT}. They have also been analyzed in optics as in \cite{Hodaei2014Parity,Ge2016_nonlinear,peng2016chiral, longhi2017unidirectional, ren2018unidirectional,madiot2024harnessing}, though the attention was not on the cubic-root-like sensitivity arising from nonlinearity.
Moreover, some works have analytically studied the behavior of {\em nonlinear} systems made of two coupled oscillators, including a limited analysis of its sensitivity \cite{Bai2023_NonlinearityES, Darcie2025_Responsitivity}. Some, as in this paper, approach these systems from a general standpoint, using a reduced-complexity model based on coupled mode theory (CMT) \cite{Zhou2016_PTSymmBreak,Assawaworrarit2017_robust, Wang2019_Chiral,Bai2023_NonlinearEP,Bai2023_NonlinearityES,Bai2024_Observations}. Other works, such as \cite{Moncada2025sensorsOscill, Moncada2025FDanalysisOscillEPD, Moncada2026_Hysterisis}, take a more direct circuit-based approach. 

In this paper, we expand the findings in \cite{Kazemi2022_CTLPTSymm, Nikzamir2022_HighlySensitive} by fully evaluating the system's sensitivity near its "degenerate" oscillation frequencies through incorporating nonlinear dynamics using CMT. To do this, we expand on an approach used for such a circuit pioneered by \cite{Zhou2016_PTSymmBreak} and developed by \cite{Assawaworrarit2017_robust, Wang2019_Chiral,Bai2023_NonlinearEP,Darcie2025_Responsitivity,Bai2023_NonlinearityES,Bai2024_Observations}, through adding a detailed sensitivity and saturated-gain analysis, an improved gain model, and energy analysis, and then apply this approach to two coupled RLC circuits. This paper presents the most comprehensive analysis of this system to date, investigating {\em degeneracies of order two and three} occurring at the saturated steady state. In particular, such degeneracies associated with heightened sensitivities are demonstrated analytically. We define SS$\mathcal{PT}$ symmetry as the regime where $\mathcal{PT}$ symmetry is imposed only after reaching a {\em saturated steady-state gain} $g_{\rm s}$ that is symmetric with respect to losses. This condition is related to operating near or at the third-order SS EPD and is also associated with bistable solutions.

The paper is organized as follows: Sec.~\ref{sec:GeneralNLHamiltonian} analyzes the steady-state oscillation frequencies and saturated gain of the nonlinear system; Sec.~\ref{sec:Sensitivity} derives the sensitivity of the SS frequencies to perturbations of parameters around the third order SS-EPD solution of the system and also around the second-order SS-EPD, and determines the best operative conditions for sensitivity; Sec.~\ref{sec:SensitivityDesign} provides an analysis for realistic implementations of enhanced sensitivity; Sec.\ref{sec:PhysicalProb} incorporates a specific gain model into the analysis to study energy and stability of the previously found steady-state solutions; and Secs.~\ref{sec:Inductive} and ~\ref{sec:Capacitive} apply this nonlinear SS-CMT analysis to two types of coupled RLC circuits analyzed in \cite{Nikzamir2022_HighlySensitive}, highlighting the method's limitations for real circuits. 
\section{General Nonlinear Approach} \label{sec:GeneralNLHamiltonian}
To study the general dynamics of this nonlinear system of coupled oscillators, we focus on its steady-state behavior, which occurs when the active element's nonlinear gain saturates to a constant value. We model energy in the oscillator using CMT, an approximate approach often applied to coupled electromagnetic oscillators such as coupled RLC circuits \cite{Assawaworrarit2017_robust, Zhou2016_PTSymmBreak,Mohseni2025_WPT} and coupled modes in waveguides and resonators \cite{Hardy1985_CMTWaveguides, Fan2003_FanoRes,Elnaggar2015_CMTResonators,Moskalev2025_CMTAnisoRR}. This modeling framework is analogous to the Hamiltonian-based approach used for quantum mechanical systems \cite{Milburn2015_linearEPD, Bai2023_NonlinearEP}, and in the literature both are often used interchangeably to formulate the general dynamics of any oscillator \cite{Mortensen2018_FluctuationsPTSymm,Wu2022_generalized}.

An important note in this section is that this saturated gain value $g_{\rm s}$ is independent of the nonlinear gain function, $g(|a_1|)$, though it must lie within the function's range \cite{Assawaworrarit2017_robust}.
Because of this, and for the sake of generality, we do not define the nonlinear gain function in this section. In Sec.~\ref{sec:PhysicalProb}, we will explore the physical implications of a specific $g(|a_1|)$ on the system.
\subsection{Nonlinear coupled equations and steady-state regimes} \label{sec:NonlinearCMT}
The oscillator under study, as depicted in Fig.~\ref{fig:CoupledOsc}, consists of two coupled resonators resonating at $\omega_2$ and $\omega_1$,  $\kappa$ is the (reciprocal) coupling coefficient \cite{Haus1984_WavesFields}, $\gamma$ is the loss in the second resonator, and $g(|a_1|)$ describes the nonlinear gain in the first resonator.
The nonlinear dynamics of this system, formulated using the CMT approach and considering positive frequencies, are given by
%
%
\begin{eqnarray} \label{eq:HE1}
\frac{da_1}{dt} = \left[ j\omega_1 + g(|a_1|) \right] a_1 - j\kappa a_2,
\\
\label{eq:HE2}
\frac{da_2}{dt} = -j\kappa a_1 + (j\omega_2 - \gamma) a_2,
\end{eqnarray}
where $a_i$ represents the state of the $i$th resonator, and $|a_i|^2$ is the energy stored in that resonator.

In order to determine the SS frequency at which this system oscillates after reaching saturation, we study its steady state, where the system's dynamics are time-invariant. This time invariance requires the nonlinear gain to converge to a constant value $g_{\rm s}$. At steady state, the oscillators synchronize such that $a_i =\tilde{a}_i e^{j\omega t}$, where the energy in each oscillator has converged to its steady-state energy $|\tilde{a}_i|^2$. This simplifies (\ref{eq:HE1}) and (\ref{eq:HE2}) to the {\em steady-state eigenfrequency problem},   
\begin{equation} \label{eq:HEMatrix}
\omega \begin{bmatrix} \tilde{a}_1  \\ \tilde{a}_2 \end{bmatrix} = 
\begin{bmatrix}
\omega_1 - j g_{\rm s} & -\kappa  \\
-\kappa & \omega_2 + j\gamma
\end{bmatrix}
\begin{bmatrix} \tilde{a}_1  \\ \tilde{a}_2 \end{bmatrix}.
\end{equation}
\noindent The eigenfrequencies are obtained by solving the characteristic equation,
\begin{equation} \label{eq:Characteristic}
\begin{split}
  & f(\omega) = (\omega_1 - \omega)(\omega_2 - \omega) + g_{\rm s} \gamma - \kappa^2 + \\
  &  \ \ \ \ \ \ \ \ \ \ \ \ \ \ \ \ \ j[\gamma(\omega_1 - \omega) - g_{\rm s} (\omega_2 - \omega)]=0.
\end{split}
\end{equation}
\noindent The two solutions are
\begin{equation} \label{eq:OmegaSol}
\omega = \frac{\omega_1+\omega_2}{2} +j\frac{\gamma - g_{\rm s}}{2}
\pm\frac{1}{2} \omega_{\Delta},   
\end{equation}
\noindent with 
\begin{equation} \label{eq:omega_Delta}
\omega_{\Delta} =\sqrt{4\kappa^2 + \left[(\omega_1-\omega_2)-   j(\gamma + g_{\rm s})\right]^2}.
\end{equation}
\noindent A degenerate solution occurs when $\omega_{\Delta}=0$.  
 
In active systems where the signal grows, the gain value $g(|a_1|)$ saturates to a value $g_{\rm s}$ associated with an SS {\em real-valued} self-oscillation frequency. Both the SS real-valued $\omega$ and saturated gain $g_{\rm s}$ are initially unknown; however, knowing one is enough to determine the other. Therefore,  through the  imposition that $\omega$ is purely real, all parameters $\omega, \omega_1, \omega_2, g_{\rm s}, \gamma$ and $\kappa$ must be real values, allowing (\ref{eq:Characteristic}) to be separated into its real and imaginary parts, 
%
%
%
\begin{eqnarray} \label{eq:ReCharacteristic}
\text{Re}[f] = (\omega_1 - \omega)(\omega_2 - \omega) + g_{\rm s} \gamma - \kappa^2 = 0,
\\
\label{eq:ImCharacteristic}
\text{Im}[f] = \gamma(\omega_1 - \omega) - g_{\rm s}(\omega_2 - \omega) = 0.
\end{eqnarray}
\noindent From these equations, we determine the SS regime, i.e., the pair SS angular frequency $\omega$ and the saturated gain value $g_{\rm s}$. The value of $g_{\rm s}$ is independent of the specific choice of the active gain's nonlinear function $g(|a_1|)$, although it must lie within the active device's gain range.

Certain insights can be drawn from (\ref{eq:ReCharacteristic}) and (\ref{eq:ImCharacteristic}) when the system parameters exhibit symmetries. From (\ref{eq:ImCharacteristic}), we infer that $\omega$ cannot equal $\omega_1$ or $\omega_2$ without being equal to both, $\omega = \omega_1 = \omega_2$. We also infer from (\ref{eq:ImCharacteristic}) that if $g_{\rm s} = \gamma$, a real steady-state oscillation frequency $\omega$ exists only if $\omega_1 = \omega_2$ (a condition that we call "steady-state (SS)$\mathcal{PT}$-symmetry"). The reverse is not necessarily true, meaning that if $\omega_1 = \omega_2$ then $g_{\rm s}$ is not necessarily equal to $\gamma$. The SS regime, found from (\ref{eq:ReCharacteristic}) and (\ref{eq:ImCharacteristic}), will be analyzed in the following sections.

\subsection{Strongly and weakly coupled regimes of the system under symmetry ($\omega_1=\omega_2$)} \label{sec:SymmetricProblem}

The most commonly studied state of this system occurs when $\omega_1=\omega_2$ is imposed. This space is divided into two regimes based on the strength of the coupling compared to the losses: the strongly coupled regime and the weakly coupled regime \cite{Hao2023_frequency,Wu2022_generalized,Guo2024_levelPinning}. 

\begin{figure}[t]
\begin{centering}
\includegraphics[width=3.5in]{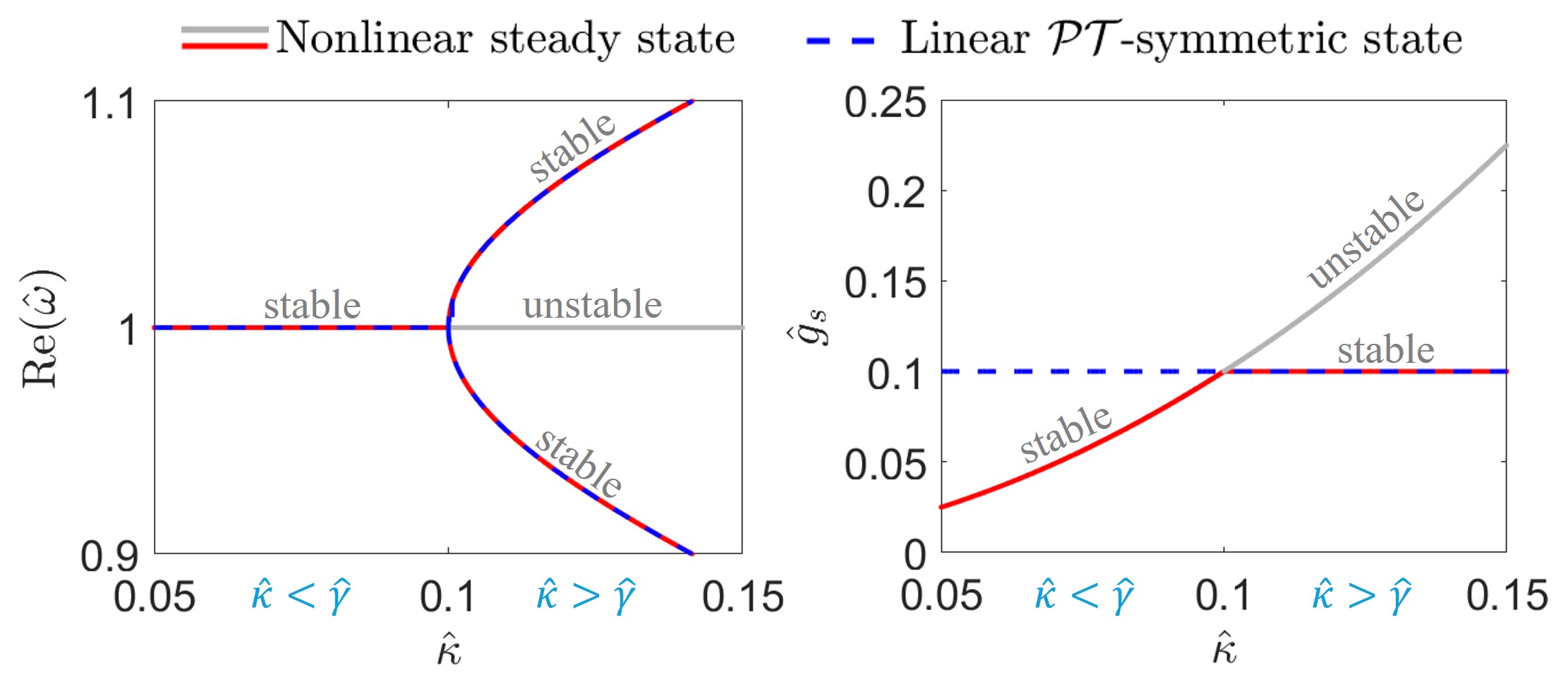}
\par\end{centering}
\caption{The saturated steady-state solution pair $\omega,g_{\rm s}$ of the symmetric system ($\omega_2=\omega_1$) from (\ref{eq:ReCharacteristicSimplified}) and (\ref{eq:ImCharacteristicSimplified}) compared against the linear solutions of the $\mathcal{PT}$ system with equivalent gain and loss ($g=\gamma$). Both solutions are plotted around the degenerate point, $\kappa=\gamma$, varying $\hat{\kappa}$ and assuming that $\hat{\gamma}=0.1$, with the hat $\hat{\ }$ denoting a normalization to $\omega_1$. 
For the nonlinear steady-state case, red and gray lines indicate stable and unstable steady-state solutions, respectively.
\label{fig:SymmetricSolutions}}
\end{figure}

Applying the symmetry $\omega_1 = \omega_2 = \omega_0$ to (\ref{eq:OmegaSol}), we find
\begin{equation} \label{eq:SimSolution}
\omega = \omega_0 +j\frac{\gamma - g_{\rm s}}{2} \pm\frac{1}{2}\sqrt{4\kappa^2 - \left(\gamma + g_{\rm s}\right)^2},
\end{equation}
and (\ref{eq:ReCharacteristic}) and (\ref{eq:ImCharacteristic}) simplify to
%
%
\begin{eqnarray} \label{eq:ReCharacteristicSimplified}
\text{Re}[f] = (\omega_0 - \omega)^2 + g_{\rm s} \gamma - \kappa^2 = 0,
\\
\label{eq:ImCharacteristicSimplified}
\text{Im}[f] = (\gamma-g_{\rm s})(\omega_0 - \omega)= 0.
\end{eqnarray}

From (\ref{eq:ImCharacteristicSimplified}), there are two distinct conditions that allow $\omega$ to be real (a necessary property for an SS oscillation frequency). The first condition is when $g_{\rm s}=\gamma$. Substituting $g_{\rm s}=\gamma$ into (\ref{eq:ReCharacteristicSimplified}) yields two corresponding steady-state oscillation angular frequencies, $\omega=\omega_0\pm \sqrt{\kappa^2-\gamma^2}$. This solution is valid only in the strongly coupled regime, $\kappa \ge \gamma$. Since both oscillation frequencies share the same saturated gain $g_{\rm s}$, this condition corresponds to a saturated or SS$\mathcal{PT}$-symmetric regime. The second condition is when $\omega = \omega_0$, and it is found in both the strongly-coupled regime and in the weakly-coupled regime ($\kappa < \gamma$). Using (\ref{eq:ReCharacteristicSimplified}), the corresponding saturated gain is $g_{\rm s} = \kappa^2/\gamma$. Unlike the first condition, there are no bounds imposed on this frequency solution existing in both regimes; however, it is only stable in the weakly coupled regime, as indicated by change of the line colors from red to gray of this steady-state pair in Fig.~\ref{fig:SymmetricSolutions}, where the hat $\hat{}$ symbol denotes normalization with respect to $\omega_1$. Stability analysis details are discussed in Sec.~\ref{sec:Stability}.


The steady-state ($\omega$, $g_{\rm s}$) pair found in the symmetric condition is plotted in Fig.~\ref{fig:SymmetricSolutions}, varying coupling. Notably, these purely real-valued steady-state frequencies closely resemble the real part of the eigenfrequency solutions of a linear $\mathcal{PT}$-symmetric system (i.e., with symmetric gain and loss). The steady-state and linear regimes share the same degenerate $\omega=\omega_1$ solution (i.e., $\hat{\omega}=1$) occurring at $g_{\rm s} = \gamma = \kappa$. In a linear system, this marks a Hopf bifurcation \cite[pp.~251-257]{Strogatz2000_Chaos,Moncada2025FDanalysisOscillEPD}, whereas in this nonlinear system it corresponds to a pitchfork bifurcation, i.e., when one branch splits into three branches \cite[pp.~56-59]{Strogatz2000_Chaos}.

\subsection{Steady-state frequency} \label{sec:Steady-State Freq}

\begin{figure}[t]
\begin{centering}
\includegraphics[width=3.5in]{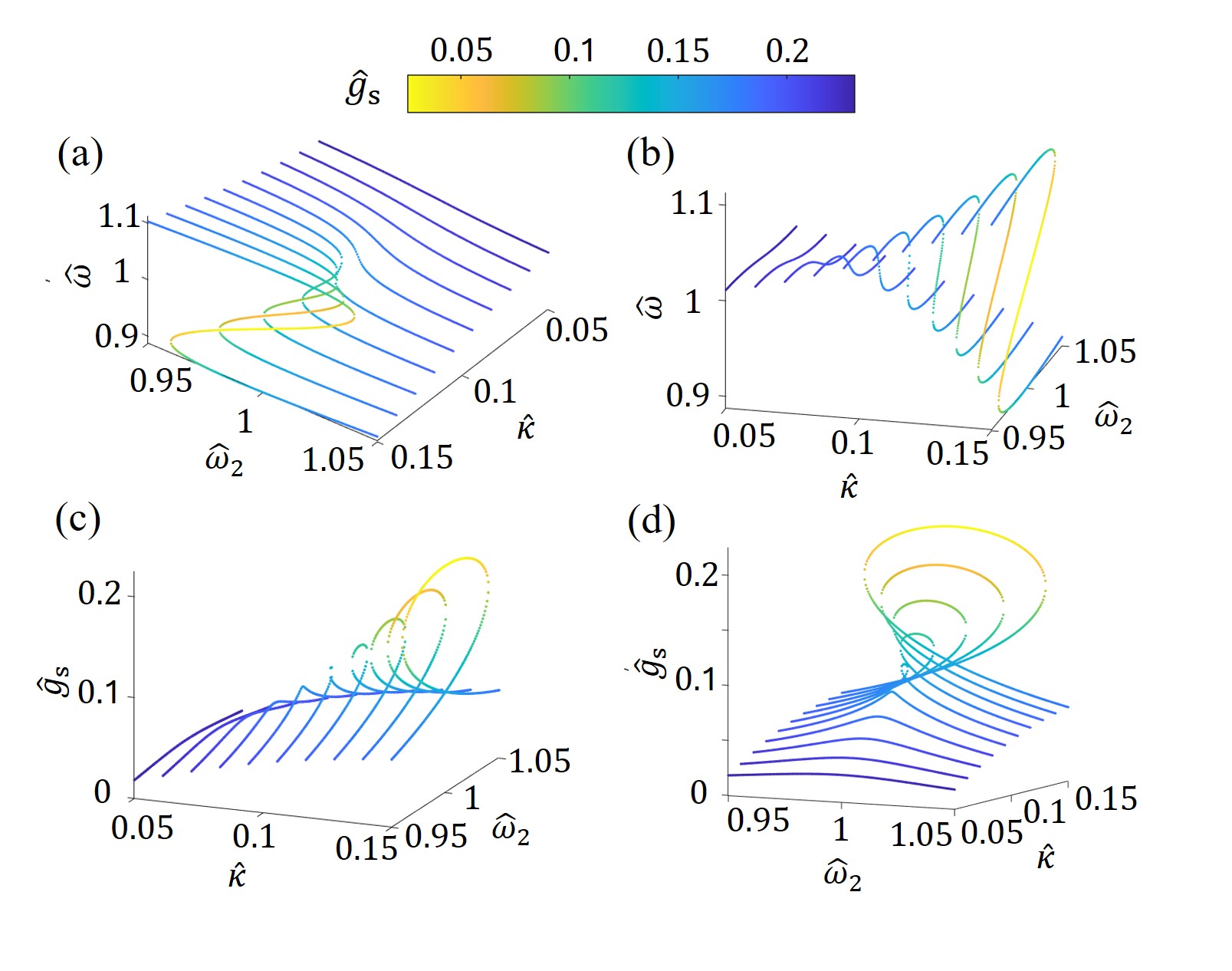}
\par\end{centering}
\caption{The three-dimensional steady-state solution space of $\hat{\omega}$ (a)-(b) (cusp catastrophe like geometry varying the two parameters) and $\hat{g}_s$ (c)-(d) varying $\hat{\omega}_2$, and $\hat{\kappa}$ around the third-order degenerate solution $\hat{\omega}_2=1$ and $\hat{\kappa}=\hat{\gamma}=0.1$. Both solution spaces are recorded from two different angles for better visualization, with the $\hat{\ }$ symbol denoting a normalization of the parameters and solutions to $\omega_1$. The colormap connects the steady-state pair between (a)-(b) and (c)-(d), indicating that when there are three solutions, the middle $\hat{\omega}$ will have the largest $\hat{g}_s$ value. As seen in (a)-(b), there is an inherent anti-symmetry across $\hat{\omega}_2=1$, which leads the system to have chiral dynamics \cite{Wang2019_Chiral}.}
\label{fig:threeD}
\end{figure}

Moving beyond the symmetric condition, the SS regime is determined by treating (\ref{eq:ReCharacteristic}) and (\ref{eq:ImCharacteristic}) as a system of equations with two unknowns, $\omega$ and $g_\mathrm{s}$. Solving for the SS oscillation frequency leads to a {\em cubic} equation, $p(\omega)=0$, where 
\begin{equation}
\label{eq:ReCharacteristicNOg}
p(\omega)=\left(\omega - \omega_1\right)\left(\omega - \omega_2\right)^2 + \gamma^2 \left(\omega - \omega_1\right) - \kappa^2 \left(\omega - \omega_2\right),
\end{equation}
that is rewritten as
\begin{equation} \label{eq:cubic}
p(\omega) = \omega^3 + b_2\omega^2 + b_1\omega + b_0.
\end{equation}
The coefficients are $b_2 = -\omega_1 - 2\omega_2$, $b_1 = \omega_2^2 + 2\omega_1\omega_2 + \gamma^2 - \kappa^2$, and $b_0 = -\omega_1\omega_2^2 - \gamma^2\omega_1 + \kappa^2\omega_2$. Each real $\omega$-solution of $p(\omega)=0$ is directly associated with a real-valued $g_{\rm s}$, which together are considered a unique steady-state solution of the system.

\begin{figure}[!t]
\begin{centering}
\includegraphics[width=3.45in]{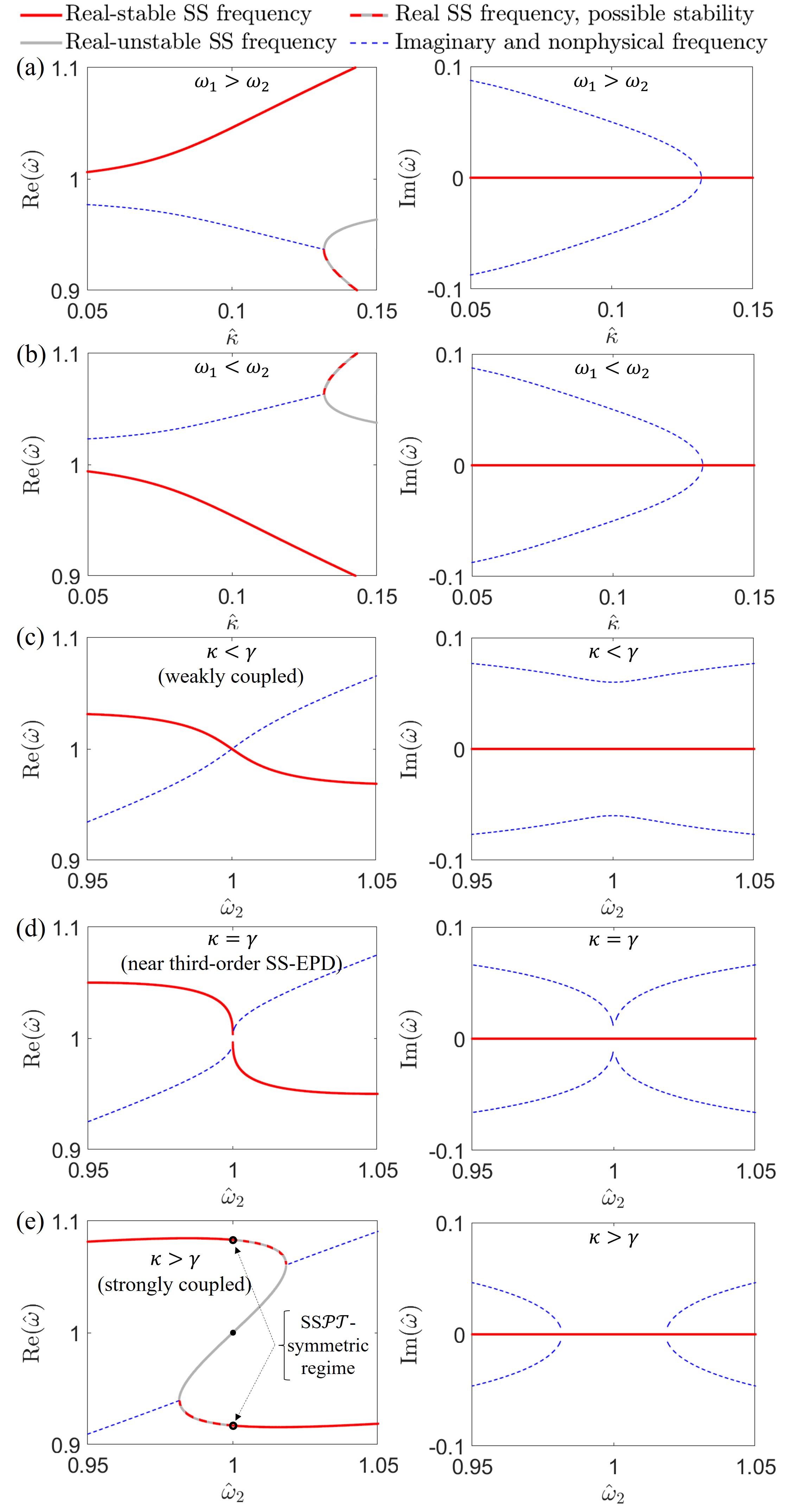}
\par\end{centering}
\caption{Steady-state oscillation frequency $\hat{\omega}$ (real-valued solutions of $p(\omega)=0$), shown in red and gray, plotted varying $\hat{\kappa}$ and $\hat{\omega}_2$ (the complex blue-dotted branches are shown for a better understanding of the solutions). Red, gray, and dashed-red colors denote the stability of the particular oscillation frequencies (see Sec.~\ref{sec:Stability}). The parameters for each of the plots are as follows, with the $\hat{\ }$ denoting a normalization to $\omega_1$. Varying $\hat{\kappa}$, with $\hat{\gamma}=0.1$: (a) $\hat{\omega}_2=0.98$; (b) $\hat{\omega}_2=1.02$. Varying $\hat{\omega}_2$, with $\hat{\gamma}=0.1$: (c) $\hat{\kappa}=0.08$; (d)  $\hat{\kappa}=\hat{\gamma}=0.1$; and (e) $\hat{\kappa}=0.13$.} 
\label{fig:ToySol}
\end{figure}

The SS oscillation frequency may differ significantly from the eigenfrequency of the corresponding linear system with small-signal gain \cite{Nikzamir2022_HighlySensitive}. Unlike the linear system, there are regions of either {\em one} SS or {\em three} SS frequencies as shown in Figs.~\ref{fig:threeD} and \ref{fig:ToySol}. These regions are separated by either double- or triple-order degenerate solutions of $p(\omega)=0$. This behavior is illustrated in the three-dimensional plots of the steady-state pairs shown in Fig.~\ref{fig:threeD}. For additional clarity, Fig.~\ref{fig:ToySol} presents the solution space of $p(\omega)=0$, whose real solutions correspond to the two-dimensional cuts of the steady-state real $\omega$-solution space depicted in Fig.~\ref{fig:threeD}.


To determine the regions supporting one or three SS frequencies and the degenerate conditions, we analyze the first derivative of the cubic polynomial, $p'(\omega) = 3\omega^2+2b_2\omega+b_1$. Local maximum and minimum of the cubic polynomial $p(\omega)$ are determined by $p'(\omega) = 0$, which occurs, respectively, at the two points
\begin{equation}
     \omega_{\substack{\mathrm{max} \\ \mathrm{min}
    }} =\frac{1}{3} \left( \omega_1 +2\omega_2 \mp \sqrt{h} \right),
\end{equation}
\noindent when $h\ge 0$, with
\begin{equation}
    h = (\omega_1-\omega_2)^2-3(\gamma ^2 -\kappa^2).
\end{equation}
%
When $h=0$, the two points coincide and $p(\omega)$ has a stationary inflection point at $\omega_\mathrm{i} = (\omega_1+2\omega_2)/3$, because both $p''(\omega_\mathrm{i})=0$ and $p'(\omega_\mathrm{i})=0$ are satisfied. 

The oscillation frequency solutions are categorized into four distinct cases. {\em Second order degenerate} oscillation frequencies occur when  $\omega_{\mathrm{max}}$ and $\omega_{\mathrm{min}}$ of the extrema exist and either $p(\omega_{\mathrm{max}})=0$ or $p(\omega_{\mathrm{min}})=0$, as discussed in Case 2 below. The {\em third order degenerate} oscillation frequency occurs when $\omega_\mathrm{i}$ exists and satisfies $p(\omega_\mathrm{i})=0$, which is discussed in Case 3 below.

{\em \textbf{Case 1}: Region of parameter space with one nondegenerate oscillation frequency}.

The SS frequency is determined to be within this region if $h < 0$ (i.e., $p(\omega)$ has no local extrema), or if $h > 0$ and either $p(\omega_{\mathrm{max}})<0$ or $p(\omega_{\mathrm{min}})>0$. This region is delineated in the plots when there exists only a single real $\omega$ for a given set of parameters. All plots in Fig.~\ref{fig:ToySol} contain sets of parameters within this region, though Fig.~\ref{fig:ToySol}(c) is the only plot where all sets of parameters are within this region.

{\em \textbf{Case 2}: Second-order SS degenerate oscillation frequency.}

A second-order SS-EPD occurs when there exists an SS degenerate oscillation frequency, corresponding to a saddle-node bifurcation \cite{Palacios2022} separating the two regions of the parameter space. This degeneracy occurs when both $p(\omega)=0$ and  $p^\prime(\omega)=0$, and $p^{\prime\prime}(\omega) \neq 0$. In other words, when either $p(\omega_{\mathrm{max}})=0$ or $p(\omega_{\mathrm{min}})=0$. One single doubly degenerate solution is seen in both Fig.~\ref{fig:ToySol}(a) and Fig.~\ref{fig:ToySol}(b) at the point separating the complex and purely real branch solutions. It is also seen twice in Fig.~\ref{fig:ToySol}(e) at the two points separating the complex and purely real branch solutions. 

{\em \textbf{Case 3}: Third-order SS degenerate oscillation frequency.}

A third-order SS-EPD oscillation frequency of the system is a unique point in the parameter space with increased sensitivity properties. This SS degeneracy occurs when the polynomial reduces to $p(\omega) = \left(\omega - \omega_i\right)^3$ implying that also  $p^\prime(\omega)=0$ and $p^{\prime\prime}(\omega) = 0$. Thus, this degeneracy condition occurs only when $\omega_1=\omega_2$ and $\gamma=\kappa$, and the steady-state oscillation frequency is $\omega=\omega_\mathrm{i}=\omega_1=\omega_2$. Indeed, in this symmetric case, $h=3(\kappa^2 - \gamma^2)$, and the required condition $h=0$ for the third order SS-EPD implies that $\kappa=\gamma$, which is the boundary between the weak and the strong coupling regimes \cite{Mohseni2025_WPT}. This third-order degenerate oscillation frequency is seen in Fig.~\ref{fig:SymmetricSolutions} at $\hat{\kappa}=0.1$ at the point separating one and three frequency solutions (pitchfork bifurcation), and also in Fig.~\ref{fig:ToySol}(d) at $\hat{\omega}_2=1$, where the slope of the solution curve diverges.

{\em \textbf{Case 4}: Region of parameter space with three oscillation frequencies.}

The steady states are determined to be within this region if $h > 0$, $p(\omega_{\mathrm{max}})>0$, and $p(\omega_{\mathrm{min}})<0$. This region is delineated where there exist three purely real $\omega$ solutions for a given set of parameters. In the plots, this region is seen in the folded regions of Fig.~\ref{fig:threeD} and also shown in Fig.~\ref{fig:ToySol} (a),(b), and (e).

\subsection{Saturated gain} \label{sec:SaturatedGain}
\begin{figure}[t]
\begin{centering}
\includegraphics[width=3.45in]{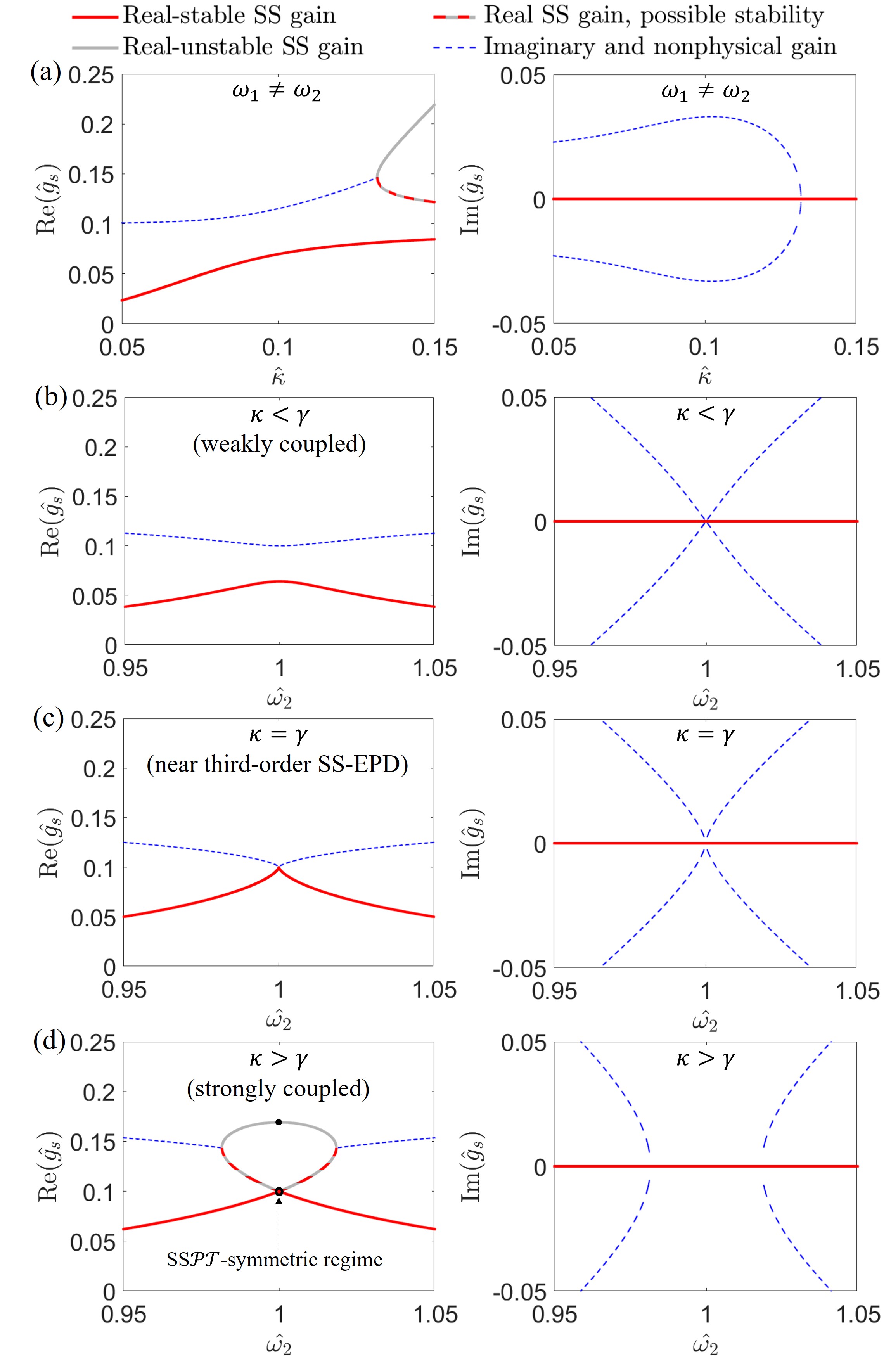}
\par\end{centering}
\caption{Saturated steady-state gain $g_{\rm s}$ from (\ref{eq:gsCubic}), in red, plotted around the third order degenerate solution varying $\kappa$ and $\omega_2$ (the complex branches, blue dotted,  are shown only for a better understanding of the solutions). Red, gray, and dashed-red colors denote the stability of the particular saturated gain values (see Sec. \ref{sec:Stability}). The parameters for each plots are as follows, with the $\hat{\ }$ denoting a normalization to $\omega_1$. (a) Varying $\hat{\kappa}$: $\hat{\gamma}=0.1$ and either $\hat{\omega}_2=0.98$ or $\hat{\omega}_2=1.02$ (symmetry across $\omega_2=\omega_1$ creates the same plots). Varying $\hat{\omega}_2$: $\hat{\gamma}=0.1$ and (b) $\hat{\kappa}=0.08$; (c) $\hat{\kappa}=\hat{\gamma}=0.1$; (d) $\hat{\kappa}=0.13$.}
\label{fig:ToySolGs}
\end{figure}

Once an SS oscillation frequency $\omega$ is determined, the associated SS gain is obtained from (\ref{eq:ImCharacteristic}), as 
\begin{equation} 
\label{eq:gs}
g_{\rm s} = \gamma \frac{\omega - \omega_1}{\omega - \omega_2}.
\end{equation}
Alternatively, the saturated gain is also found from the cubic equation,
\begin{equation} \label{eq:gsCubic}
g_{\rm s}^3 - \left(2\gamma + \frac{\kappa^2}{\gamma}\right) g_{\rm s}^2 + \left( \gamma^2 + 2\kappa^2 + (\omega_2 - \omega_1)^2 \right) g_{\rm s} - \gamma\kappa^2= 0.
\end{equation}
Three-dimensional steady-state plots of this saturated gain are provided in Fig.~\ref{fig:threeD}, varying the two parameters $\omega_2$ and $\kappa$. For clarity, Fig.~\ref{fig:ToySolGs} shows selected two-dimensional slices of the saturated-gain space in Fig.~\ref{fig:threeD}, calculated using (\ref{eq:gsCubic}).

Studying the saturated-gain solutions of the cubic polynomial, we find the same groups as for the steady-state oscillation frequency solutions: one, two degenerate and one separate, three degenerate, or three separate real saturated gain values. We also observe that the $g_{\rm s}$ values are completely symmetric over the sign of $(\omega_2-\omega_1)$ (or equivalently the sign of $ \hat{\omega}_2-1$, where $\hat{\omega}_2 = \omega_2/\omega_1$). This symmetry in the $g_{\rm s}$ values is connected to an anti-symmetry (odd function symmetry) across $\omega_2=\omega_1$ in the associated steady-state frequencies, as seen in Fig.~\ref{fig:threeD}. 

Another useful connection between the saturated gain and steady-state frequency is obtained by inserting (\ref{eq:ReCharacteristicNOg}) into (\ref{eq:gs}), leading to 
\begin{equation} 
\label{eq:gs-omega2}
g_{\rm s} =  \frac{\gamma \kappa^2}{\left(\omega - \omega_2\right)^2 + \gamma^2 }.
\end{equation}

In general, $g_{\rm s} \leq \kappa^2/\gamma$, with the largest value $g_{\rm s}=\kappa^2/\gamma$ occurring when $\omega=\omega_1=\omega_2$ in both the strongly and weakly-coupled regimes. In the strongly-coupled regime, exemplified in Fig.~\ref{fig:ToySolGs}(d), the maximum gain graphically corresponds to the steady-state frequency in the middle region of the three real oscillation frequencies shown in Fig.~\ref{fig:ToySol}(e), which is, however, unstable as discussed in Sec.~\ref{sec:Stability}. The other two SS oscillation frequencies are either both stable (and never satisfying $\omega \ne\omega_1$), forming the bistable regime, or at least one is stable. When $\omega_2=\omega_1$, the two SS oscillation frequencies have the same saturated gain $g_{\rm s}$. This SS gain degeneracy is not an SS-EPD and is associated with two SS$\mathcal{PT}$-symmetric regimes because $g_{\rm s}=\gamma$ for both SS frequencies as seen in Fig.~\ref{fig:SymmetricSolutions} and from (\ref{eq:ImCharacteristicSimplified}).

In the weakly coupled regime, exemplified in Fig.~\ref{fig:ToySolGs}(b), the maximum gain graphically corresponds to the SS oscillation frequency shown in Fig.~\ref{fig:ToySol}(c), which is stable. However, when $\kappa=\gamma$, the largest value the saturated gain can assume is $g_{\rm s}=\gamma$ which occurs at the third-order SS-EPD, as seen in  Fig.~\ref{fig:ToySolGs}(c). 

The saturated gain directly impacts the physical system in terms of the stability of the steady-state solutions, the energy contained in each oscillator, and which stable frequency the system initially tends to oscillates at. These topics are covered in Sec.~\ref{sec:PhysicalProb}.

\section{Exceptional sensitivity of the oscillation frequency to perturbations} \label{sec:Sensitivity}

One of the celebrated properties of EPDs is their sensitivity to perturbations \cite{Kato1966_perturbation,Binkowski2024_WiersigEPSensitivity,Kazemi2019_LinearPTEPD}. In a {\em linear} system, the variation  $\Delta\omega=\omega -\omega_\mathrm{0}$ of the eigenfrequency $\omega$ from the EPD at $\omega_\mathrm{0}$, shows "exceptional" sensitivity to a small perturbation $\Delta X=X-X_0$ of a system parameter $X$ near the EPD parameter $X_0$. This sensitivity is approximated by the first term of the Puiseux fractional power expansion \cite{welters2011explicit},  
\begin{equation} 
\label{eq:Puiseux1}
\Delta \omega \propto \sqrt{\Delta X}.
\end{equation}

In the case of {\em nonlinear saturable gain}, the sensitivity of the steady-state frequency of oscillation $\omega$ to a perturbation $\Delta X$ may differ from that of a linear system. For this case, the variation of the oscillation frequency $\Delta \omega$ as a function of a parameter perturbation $\Delta X$ relies on the properties of the oscillation frequency $\omega(X)$ determined by (\ref{eq:ReCharacteristic}) and (\ref{eq:ImCharacteristic}). In other words, we look at the variation $\Delta \omega$ found from the zeros of the polynomial equation $p(\omega)=0$ when a parameter $X$ in its coefficients is perturbed. It is convenient here to generalize the notation in (\ref{eq:cubic}) by considering the same polynomial as a function of two variables, $p(\omega,X)=0$, while also assuming that the solution $\omega$ of such polynomial is a function of $X$. In this section, we study the sensitivity of the steady-state oscillation frequencies $\omega$ to a perturbation $X$, namely, we evaluate  $d\omega/dX$ in the neighborhood of a given $\omega_0, X_0$, which still satisfy  $p(\omega,X)=0$ and $p(\omega_0,X_0)=0$. In this context, $\omega_0$ represents the region of operation that is perturbed while observing the sensitivity of the oscillation frequency to such perturbation. In this analysis, we will assume different values of the regime parameter $X_0$ associated with the third and the second order SS-EPD, as well as points near them.
We focus on identifying the cases that have extremely high (i.e., exceptional) sensitivity. 

The sensitivity $d \omega/dX$ of the function $\omega(X)$ is obtained by applying the implicit function theorem to $p(\omega,X)=0$:
\begin{equation}
 \label{eq:ImplicitFnTheorem}
\frac{d \omega(X)}{dX} = -\frac{\partial p / \partial X}{\partial p/\partial \omega}.
\end{equation}
Here, $\partial p / \partial \omega = 3\omega^2 + 2b_2\omega + b_1=p^{\prime}(\omega)$, where, for convenience, we continue to use the notation $p^{\prime}(\omega)$ adopted previously. 
Clearly, the oscillation frequency $\omega$ is infinitely sensitive when $p^{\prime}(\omega)=0$, and this special condition is encountered in two distinct degeneracies, of order two and three, each characterized by distinct dynamical behavior. Therefore, as discussed in Sec.~\ref{sec:Steady-State Freq}, the system experiences the highest (i.e., exceptional) sensitivity at the operating oscillation frequencies
\begin{equation} \label{eq:ExceptionalSensitive}
\omega(X_0)=\omega_{\mathrm{max}}   \;\;\;\;\;\;
\textrm{and}  \;\;\;\;\;\; \omega(X_0)=\omega_{\mathrm{min}}
\end{equation}
when $h\ge 0$, as well as at the higher-order degenerate point $\omega(X_0)=\omega_{\mathrm{max}}=\omega_{\mathrm{min}}$ when $h=0$.

\begin{figure}[t]
\begin{centering}
\includegraphics[width=3.5in]{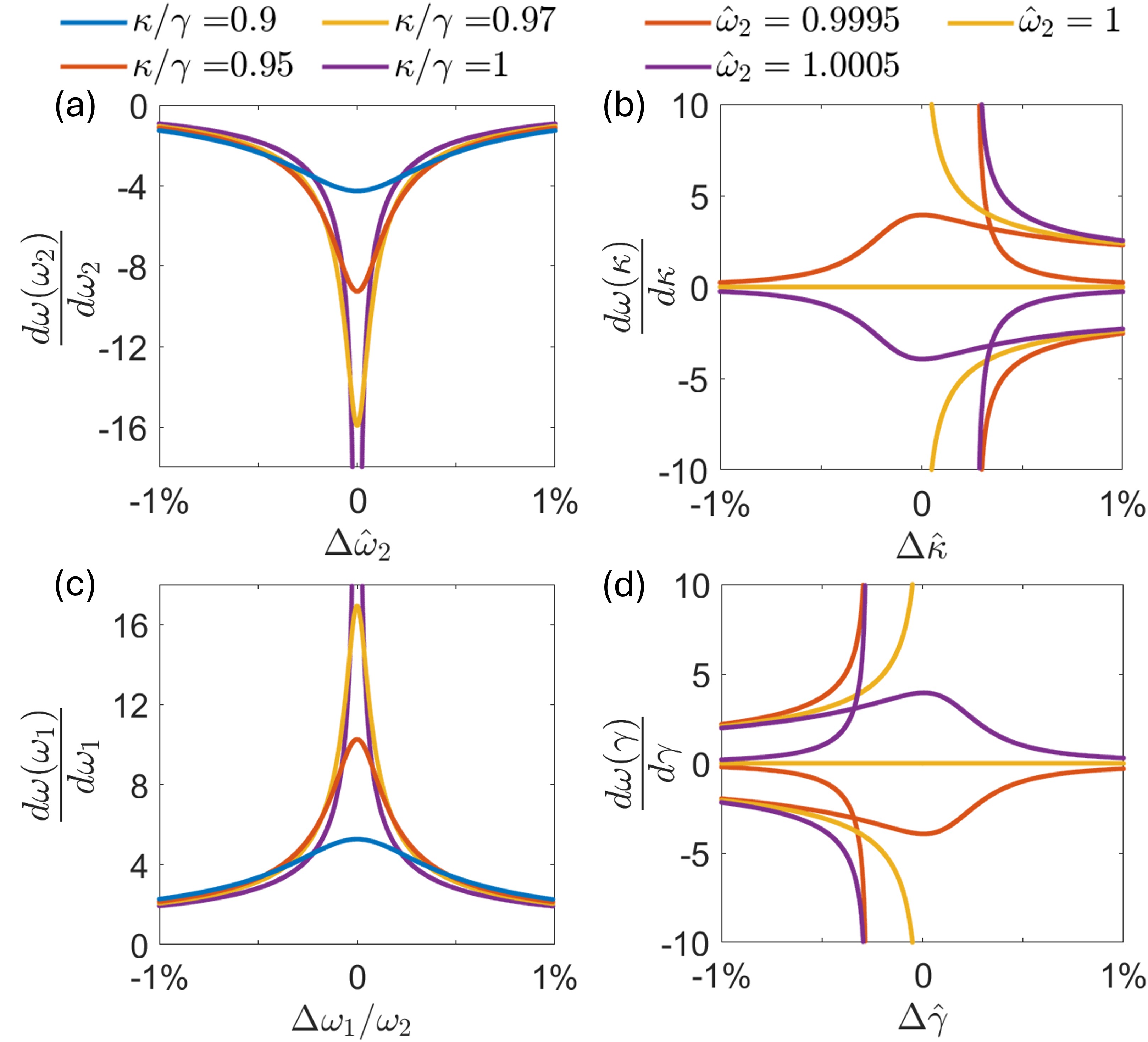}
\par\end{centering}
\caption{Plots of the sensitivity of the SS oscillation frequency $\omega(X)$ to variations in $X=\omega_2, \kappa,\omega_1,\gamma$, calculated using (\ref{eq:ImplicitFnTheorem}), including multiple curves in the neighborhood of the third-order SS-EPD. 
In each plot, the $\hat{\ }$ denotes normalization to $\omega_1$. 
The separate plots are: (a) $d\omega/d\omega_2$ with $\hat\gamma=0.1$ and $\Delta \omega_2 = \omega_2-\omega_1$; (b) $d\omega/d\kappa$ with $\hat\gamma=0.1$ and $\Delta \kappa = \kappa-\gamma$; (c) $d\omega/d\omega_1$ with $\gamma=0.1\omega_2$, holding $\omega_2$ constant, and $\Delta \omega_1 = \omega_1-\omega_2$; and (d) $d\omega/d\gamma$ with $\hat\kappa=0.1$ and $\Delta \gamma = \gamma-\kappa$.} 
\label{fig:Derivative}
\end{figure}

We include plots of the derivatives obtained using (\ref{eq:ImplicitFnTheorem}) in Fig.~\ref{fig:Derivative} to aid our description of the sensitivity to each parameter. The exceptional sensitivity is evident in these plots, which occurs when the derivatives diverge. We only include the cases with $\kappa\le\gamma$ for the derivatives $d\omega/d\omega_1$ and $d\omega/d\omega_2$, which is the ideal range for operation of an exceptional-based sensor where sensitivity is increased, as this avoids the region of multiple solutions. 

\subsection{Exceptional cubic-root sensitivity of $\omega$ to perturbations of $\omega_2$, assuming $\gamma=\kappa$}
\label{sec:CaseCubicSensitivity}

First, we focus on the sensitivity of the SS oscillation frequency $\omega$ with respect to perturbations in $\omega_2$. 
It is visible in Fig.~\ref{fig:ToySol}(d), which also shows the third-order SS-EPD at $\omega=\omega_1$ when $\omega_2=\omega_1$. The sensitivity is illustrated in Fig.~\ref{fig:Derivative}(a). To determine the analytic expression for the sensitivity, we use  (\ref{eq:ImplicitFnTheorem}) with $X=\omega_2$, which leads to
\begin{equation} 
\label{eq:Sol1disersion equationNL}
\frac{d \omega(\omega_2)}{d\omega_2} = \frac{2 \omega^2 -2(\omega_1+\omega_2)\omega +2 \omega_1 \omega_2 - \kappa^2}{p^{\prime}(\omega)}.
\end{equation}
Infinite sensitivity occurs when operating at $\omega_2=\omega_1$, which leads the oscillation frequency to also be  $\omega=\omega_1=\omega_2$, solution of $p(\omega)=0$, before perturbing $\omega_2$.

Holding $\omega_1$ constant, we define $\Delta \omega=\omega-\omega_1$ and perturb $\omega_2$ by $\Delta \omega_2 = \omega_2-\omega_1$. When $\gamma=\kappa$, we find the exceptional sensitivity of the oscillation frequency to a perturbation in $\omega_2$ to be  
\begin{equation} 
\label{eq:SensitivityDegenr}
\Delta \omega \approx  - \kappa  ^{2/3}(\Delta \omega_2)^{1/3},
\end{equation}
through using (\ref{eq:Sol1disersion equationNL}), as detailed in Appendix \ref{Appendix:w2(w)Sensitivity}. This shows that the oscillation frequency perturbation $\Delta \omega$ exhibits cubic-root sensitivity to a perturbation in the resonance frequency of the second oscillator, $\omega_2$. The region in which this cube-root sensitivity is valid while $\Delta\omega$ is small, as seen by the agreement between the red and green curves in the rightmost plot of Fig.~\ref{fig:Sensitivity}. 

This sensitivity relation was previously calculated in \cite{Bai2023_NonlinearityES, Darcie2025_Responsitivity} through a different analysis, similar to that in Appendix \ref{Appendix:Taylor}. It is difficult to operate precisely at such point that exhibits cubic-root sensitivity, as in practice some difference may arise between the values of $\gamma$ and $\kappa$ or between $\omega_1$ and $\omega_2$. Hence, we next determine the sensitivity of these important cases.
\subsection{High sensitivity of $\omega$ to perturbations of $\omega_2$, assuming $\gamma \ne \kappa$} 
We study the sensitivity of the steady-state oscillation frequency $\omega$ to perturbations in $\omega_2$ near $\omega_2=\omega_1$, assuming that $\gamma\ne \kappa$. Two example curves of this case are shown in Fig.~\ref{fig:ToySol}(c) and Fig.~\ref{fig:ToySol}(e).
In this case, when $\omega_2=\omega_1$, $\omega=\omega_1$ is always a solution of $p(\omega)=0$ (Sec.~\ref{sec:SymmetricProblem}).

When $\omega_2=\omega_1$ and $\gamma \ne \kappa$, the denominator of (\ref{eq:Sol1disersion equationNL}) does not vanish as $p^{\prime}(\omega=\omega_1) = (\gamma^2-\kappa^2) \ne 0$, indicating that there is no degeneracy in this case. As proved in Appendix \ref{Appendix:w(w2)Sensitivity}, the sensitivity near the steady-state oscillation frequency $\omega=\omega_1=\omega_2$ is given by 

\begin{equation} 
\label{eq:sensitivity-Normal}
\Delta \omega \approx \frac{-\kappa^2}{\gamma^2-\kappa^2} \Delta \omega_2.
\end{equation}
This shows that the sensitivity is mainly linear for small $\Delta \omega_2$ (i.e., $\Delta \omega \propto \Delta \omega_2$). However, the sensitivity tends to infinity when we approach the case with $\gamma=\kappa$, which is the case treated in the previous section.

In summary, we have shown that when $\gamma=\kappa$, the sensitivity follows the cubic-root dependence in (\ref{eq:SensitivityDegenr}), showing infinite value when $\Delta \omega_2=0$, whereas when $\gamma\ne \kappa$ the sensitivity is mainly linear. As $(\gamma - \kappa) \rightarrow 0$, the linear coefficient $\kappa^2/(\gamma^2-\kappa^2)$ tends to infinity, as also shown through the derivatives in Fig.~\ref{fig:Derivative}(a). The steady-state frequency's increased sensitivity when $\gamma\approx \kappa$ is also apparent in the plots in Fig.~\ref{fig:Sensitivity}. 

It is important to note that when $\kappa>\gamma$, three SS oscillation frequencies exist (see Fig.~\ref{fig:threeD} and Fig.~\ref{fig:ToySol}(e)), and the sensitivity described by (\ref{eq:sensitivity-Normal}) is valid only for the middle of these three frequencies, which is always unstable \cite{Wang2019_Chiral}. As a result, in any experimental measurement of the system’s sensitivity, one would observe variations in one or both of the two other oscillation frequencies, as either one or both are stable steady states (see Sec.~\ref{sec:Stability}). Since these two frequencies exhibit significantly lower sensitivity to changes in $\omega_2$, systems designed to enhance sensitivity to perturbations should instead operate in the weakly coupled regime, $\kappa\le\gamma$. A detailed discussion of the benefits of operating in each regime for two capacitively coupled LC circuits is included in \cite{Moncada2026_Hysterisis}, as operation in the strongly coupled regime can lead to hysteresis loops.

\subsection{Exceptional cubic-root sensitivity of $\omega$ to perturbations of $\omega_1$, assuming $\gamma=\kappa$}
Applying the same steps used in Sec.~\ref{sec:CaseCubicSensitivity} to perturbations in $\omega_1$, while keeping $\omega_2$ constant, we find that $\omega$ has the same cube-root sensitivity around the third-order degenerate solution (i.e., $\omega \approx \omega_\mathrm{i}=\omega_1=\omega_2$ and $\gamma=\kappa$) to perturbations in $\omega_1$ near $\omega_2$. Indeed, by defining $\Delta \omega=\omega-\omega_2$, and $\Delta \omega_1 = \omega_1-\omega_2$, we obtain 
\begin{equation}
\Delta \omega \approx  \kappa  ^{2/3}(\Delta \omega_1)^{1/3}. 
\end{equation}
When $\gamma\neq\kappa$, the approximate sensitivity is linear, i.e., $\Delta \omega \propto \Delta \omega_1$. In both cases, when $\gamma=\kappa$ and $\gamma\neq\kappa$, the sensitivity of $\omega$ to perturbations in $\omega_1$ is similar to the sensitivity of $\omega$ to perturbations in $\omega_2$. However, since $p(\omega)=0$ is not symmetric with respect to $\omega_1$ and $\omega_2$, any expansion exhibits slight differences in higher-order terms. This is verified by the small difference in magnitude of the derivative between Fig. \ref{fig:Derivative}(a) and Fig. \ref{fig:Derivative}(c), seen when observed closely. An additional method verifying these sensitivity results is included in Appendix \ref{Appendix:Taylor}.
%
\subsection{Exceptional square-root sensitivity of $\omega$ to perturbations of $\kappa$ or $\gamma$, assuming $\omega_2=\omega_1$}
\label{sec:sqrt-Sensitivity}
We determine the sensitivity of the steady-state oscillation frequency $\omega$ to perturbations in $\kappa$ or $\gamma$, around the operating regime with $\omega_2=\omega_1$ and $\kappa=\gamma$. As discussed earlier, under this condition, the oscillation frequency is $\omega =\omega_2=\omega_1$, before perturbing either $\kappa$ or $\gamma$. The sensitivity is found directly from the solutions for the symmetric case ($\omega_2=\omega_1$) discussed in Sec.~\ref{sec:SymmetricProblem}, and plotted in Fig. \ref{fig:SymmetricSolutions}. 

In the weakly coupled case, $\kappa<\gamma$, the steady-state oscillation frequency is constant, $\omega=\omega_0$ (see Fig.~\ref{fig:SymmetricSolutions}), and it has neither $\kappa$ nor $\gamma$ dependency; thus, it exhibits no sensitivity. In the strongly coupled case, $\kappa > \gamma$, there are two (stable) steady-state oscillation frequencies such that $\Delta\omega = \pm\sqrt{\kappa^2-\gamma^2}$, where $\Delta \omega=\omega-\omega_0$. Thus, when studying the sensitivity of $\omega(\kappa)$ around $\kappa\approx\gamma$, we have 
\begin{equation}
\label{eq:senstok} 
\Delta \omega \approx \pm\sqrt{2\gamma (\kappa-\gamma)},
\end{equation}
showing the oscillation frequency's square-root sensitivity to $\kappa$ near $\gamma$.

Analogously, the sensitivity of $\omega(\gamma)$ is
\begin{equation}
\label{eq:senstogamma} 
\Delta \omega \approx \pm\sqrt{2\kappa (\kappa-\gamma)}
\end{equation}
for values of $\gamma<\kappa$.

It may be difficult to operate precisely at such a point that exhibits square-root sensitivity, as in practice some difference may arise between the values of $\omega_1$ and $\omega_2$. Hence, we next determine the sensitivity of this case.

\subsection{High sensitivity of $\omega$ to perturbations of $\kappa$ or $\gamma$, assuming $\omega_2\neq\omega_1$}

In finding the sensitivity of the oscillation frequency $\omega(\kappa)$ to variations in $\kappa$ when $\omega_2\neq\omega_1$, we once again use the implicit function theorem method previously applied. The plots showing the relation $\omega(\kappa)$ for this case are included in Fig.~\ref{fig:ToySol}(a) and Fig.~\ref{fig:ToySol}(b). The details of this application are covered in Appendix \ref{Appendix:w(k)Sensitivity}. The sensitivity is found to be linear for $\kappa \approx \gamma$:
\begin{equation} \label{eq:ksensitvity-Normal}
    \Delta\omega \approx -\frac{2\gamma}{2\omega_1 + \omega_2 -3\omega_0}\Delta\kappa,
\end{equation}
where $\Delta\kappa=\kappa-\gamma$, $\Delta\omega = \omega-\omega_0$ and $\omega_0$ is the real-valued oscillation frequency when $\kappa=\gamma$, i.e., $p(\omega_0)=0$, shown in Appendix \ref{Appendix:w(k)Sensitivity}. We note that when $\kappa = \gamma$ and $\omega_2\neq\omega_1$, only a single steady-state $\omega$ exists, shown in Fig.~\ref{fig:ToySol}(d), whose sensitivity to perturbations in $\kappa$ is captured by (\ref{eq:ksensitvity-Normal}).

We similarly find the sensitivity of $\omega(\gamma)$ to variations of when $\gamma$ near $\kappa$ to be approximately linear, 
\begin{equation} \label{eq:gammasensitvity-Normal}
    \Delta\omega \approx \frac{2\kappa(\omega_1-\omega_0)}{(2\omega_1 + \omega_2 -3\omega_0)(\omega_2-\omega_0)}\Delta\gamma,
\end{equation}
where $\Delta\gamma=\gamma-\kappa$.

In summary, at $\omega_2=\omega_1$, in the strongly coupled regime ($\kappa>\gamma$), the stable oscillation frequencies exhibit square-root sensitivity to perturbations of $\kappa$ or $\gamma$, when working under $\gamma \approx \kappa$. However, when $\omega_2\neq\omega_1$, the sensitivity to perturbation of $\kappa$ or $\gamma$ is mainly linear, but as $\omega_2 - \omega_1 \rightarrow 0$, around which $\omega_0$ also converges to $\omega_1$, the linear coefficients in (\ref{eq:ksensitvity-Normal}) and (\ref{eq:gammasensitvity-Normal}) tend to infinity, causing the steady-state frequency's increased sensitivity around $\kappa = \gamma$ as shown in Fig. \ref{fig:Derivative}(b) and Fig.~\ref{fig:Derivative}(d).

We also note that when $\omega_2\neq\omega_1$, the sensitivity of $\omega$ to perturbations of either $\kappa$ or $\gamma$ is not symmetric about $\kappa=\gamma$, and becomes extremely asymmetric when $\omega_2 = \omega_1$, as shown in Fig.~\ref{fig:Derivative}. Therefore, the region of operation for any sensing application involving the perturbation in $\kappa$ or $\gamma$  should be chosen accordingly. Additionally, when $\omega_2\neq\omega_1$ , the direction of variation of $\omega$, in response to perturbations of $\kappa$ and $\gamma$, depends on the relative values of $\omega_1$ and $\omega_2$. For $\omega_1>\omega_2$, the oscillation frequency $\omega$ increases with increasing $\kappa$ and decreases with increasing $\gamma$. Conversely, for $\omega_1<\omega_2$, $\omega$ decreases with increasing $\kappa$ and increases with increasing $\gamma$. This behavior is seen by studying Fig.~\ref{fig:ToySol}(a) and Fig.~\ref{fig:ToySol}(b) or by observing the sign of the derivatives of the continuous solution in Fig.~\ref{fig:Derivative}(b) and  Fig.~\ref{fig:Derivative}(d). The behavior could also be exploited in a calibration algorithm to tune the system closer to the degenerate conditions in a practical implementation.

\section{Exceptionally sensitive design} \label{sec:SensitivityDesign} 
We explore several other aspects of this system with respect to applying this exceptional sensitivity to a realistic design.

\subsection{Retrieval of perturbed parameter $\omega_2$ from oscillation frequency shift $\Delta \omega$} \label{sec:PerturbedParam} 

For the benefit of a real design, we assume to operate at an oscillatory steady-state regime with $\omega=\omega_1$ when the system is unperturbed, i.e., when  $\omega_2=\omega_1$, assuming fixed parameters $\gamma$ and $\kappa$ such that $\kappa \le \gamma$.  
Then, when a perturbation is applied to $\omega_2$, it is important that from the reading of the shift of the frequency of oscillation $\Delta \omega$ one can estimate the perturbation $\Delta \omega_2=\omega_2-\omega_1$.  This can be done by calibration, though we show here a simple formula, found in Appendix \ref{Appendix:w2(w)Sensitivity}, that provides a good approximation of the shift $\Delta \omega_2$:

\begin{equation} \label{eq:TaylorSeriesExp}
\Delta\omega_2 \approx \frac{\kappa^2 - \gamma^2}{\kappa^2}\Delta\omega - \frac{\gamma^4}{\kappa^6}\Delta\omega^3.
\end{equation}

\begin{figure}[t]
\begin{centering}
\includegraphics[width=3.45in]{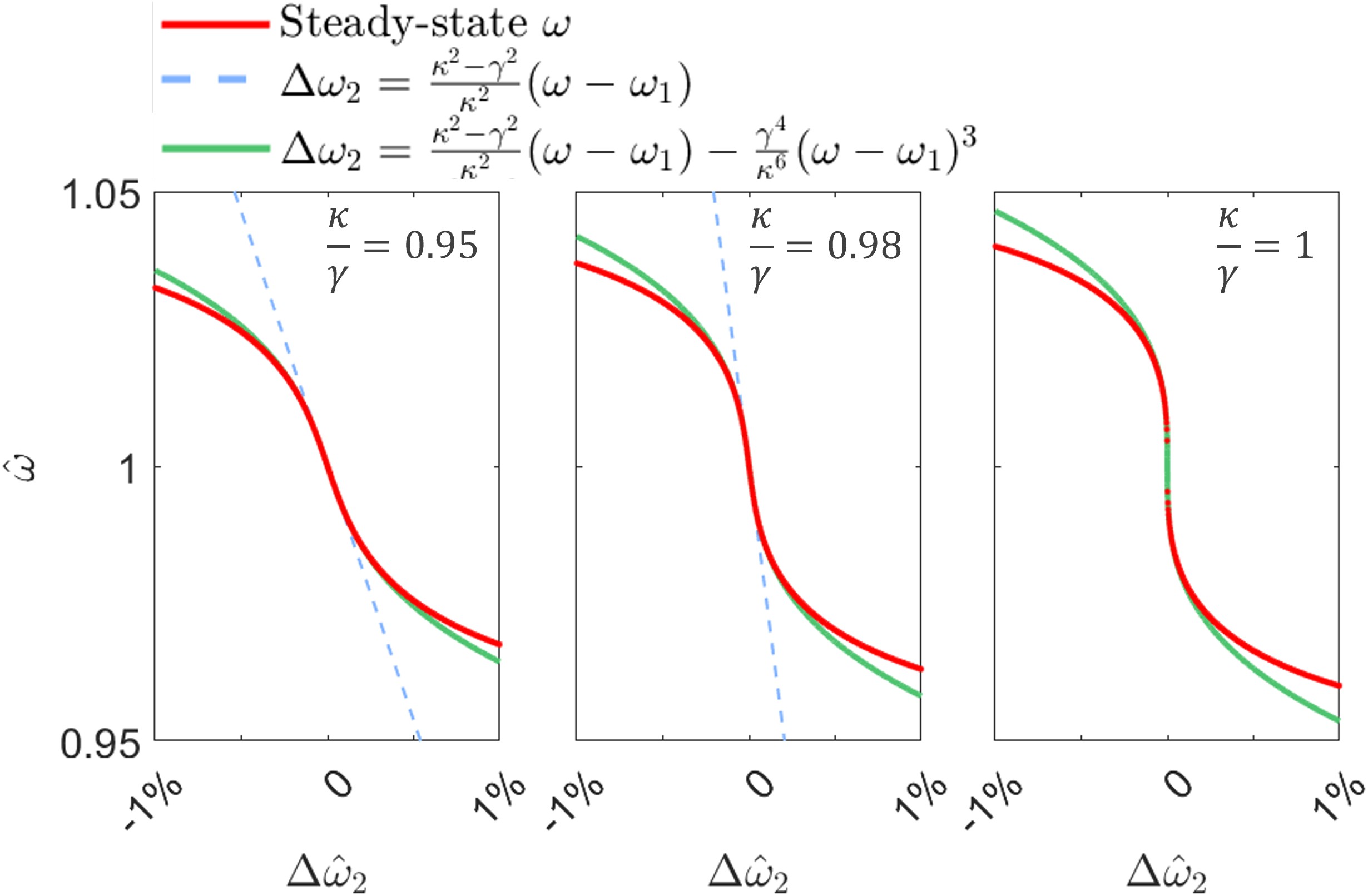}
\par\end{centering}
\caption{Plots showing the observable steady-state oscillation frequency $\omega$ caused by the percentage variation of $\omega_2$, compared to the first two terms of the Taylor series expansion for $\omega_2(\omega)$. Two are weakly coupled cases ($\kappa<\gamma$), whereas the right-most one exhibits the cubic-root sensitivity (vertical asymptote) when $\kappa=\gamma$. These three plots show the progressive decrease of the magnitude of the linear term in (\ref{eq:TaylorSeriesExp}) as $\kappa \rightarrow\gamma$, resulting in higher sensitivity to perturbations.} 
\label{fig:Sensitivity}
\end{figure}

This formula consists of the first two terms of the Taylor series expansion around $\Delta\omega=0$, and thus is applicable for small $\Delta\omega$. Including the second term of the expansion greatly increases the range of frequencies this expansion approximates, especially when $\kappa/\gamma\approx 1$ as seen in Fig.~\ref{fig:Sensitivity}.  This case represents probably the most important regime of operation because of its high sensitivity (almost like a cubic root), and that is why we provide an analytical solution. 

The retrieval of the other parameters when perturbed, by observing the oscillation frequency shift $\Delta \omega$, can also be derived analytically, or it can be inferred from the solutions of the previous section when only the first order is sufficient.

\begin{figure*} \centering
  \includegraphics[width=0.86\textwidth]{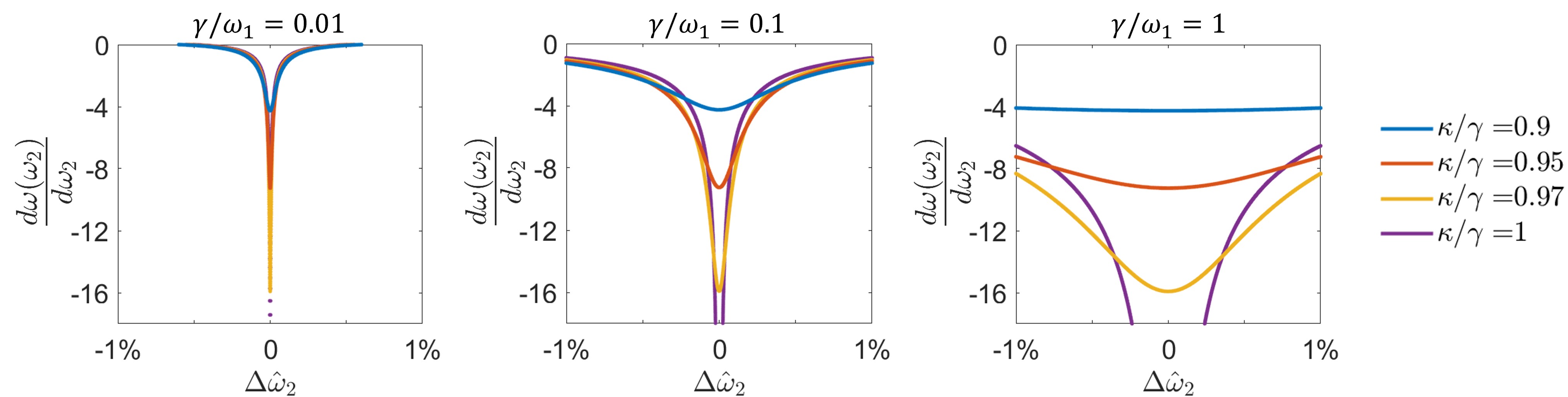}
  \caption{Plots of the sensitivity of the oscillation frequency, $d\omega/d\omega_2$, calculated using (\ref{eq:ImplicitFnTheorem}) when operating near the third-order SS-EPD (that occurs when $\omega_2=\omega_1$ and at $\gamma=\kappa$) for four values of $\kappa$ near $\gamma$, assuming three distinct cases of $\gamma/\omega_1$ ratios. This ratio directly controls the percentage of $\omega_2$ perturbations that have high sensitivity. A larger $\gamma$ value (i.e., larger losses) increases the $\omega_2$ range of high sensitivity without compromising the value of the maximum sensitivity. }
\label{fig:SensitivityRatio}
\end{figure*}

\subsection{Effect of the proportional relation between degenerate SS oscillation frequency and saturated gain.} \label{sec:ProportionalEffects} 

Thus far in this paper, when plotting the steady-state $\omega$ and $g_{\rm s}$ for sets of parameters, we have held $\gamma$ and $\omega_1$ constant, maintaining the ratio $\gamma/\omega_1 = 0.1$ ($\hat\gamma=0.1$). Maintaining this ratio biases the results to only show the exceptional variation and sensitivity of the steady-state $\omega$ and $g_{\rm s}$ around a third-order SS-EPD with the same ratio, $g_{\rm s} = 0.1 \omega$ ($\omega= \omega_1=\omega_2$, $g_{\rm s} = \gamma=\kappa$). 

Changing the ratio between $\gamma$ and $\omega_1$ affects the scale of the variation of the steady-state $\omega$ and $g_{\rm s}$ to the changes of the parameters, especially around the degenerate points. 
This is illustrated in Fig.~\ref{fig:SensitivityRatio}, which shows how different ratios between $\gamma$ and $\omega_1$ affect the proportional range of the $\omega_2$ values with increased sensitivity. Thus, a larger ratio leads to a much larger region of increased sensitivity around degenerate solutions and consequently a larger range of steady-state $\omega$ and $g_{\rm s}$ with respect to changes in $\omega_2$. Though less significant than other findings, this can have important impacts in any realistic implementation of this analysis. 
This also applies to maintaining a constant ratio between any combination of $\omega_1$ or $\omega_2$ and $\gamma$ or $\kappa$ and varying the other non-constant parameters.
%
\section{Energy balance, gain model, stability analysis, and time domain solutions} \label{sec:PhysicalProb}
We examine additional aspects of the steady-state regime, including the energy and the {\em stability} of each steady-state oscillation frequency $\omega$ and saturated gain $g_{\rm s}$. In general, these aspects depend on a gain model (in contrast to the steady-state $\omega$ and $g_{\rm s}$ that do not depend on the particular gain model parameters), and so a realistic model will be included.

\subsection{Energy balance and the steady-state eigenvectors}
We consider the energy balance at steady state and examine the saturated system's eigenvectors. In general, energy conservation is a defining feature of a closed system. This concept extends to linear open systems with the application of $\mathcal{PT}$-symmetry \cite{Bender2019_PTSymmBook}. In a $\mathcal{PT}$-symmetric system, the energy in the system is not conserved, but the energy entering and leaving the system remains balanced, so that the system's energy remains constant. The principles of energy balance can also be applied to our system at steady state, as the system converges to a constant energy value. 

The total energy stored in the two resonators is $W = |\tilde a_1|^2 + |\tilde a_2|^2$ \cite{Haus1984_WavesFields}. At steady state, the balance of energy imposes that $\frac{d}{dt}(|\tilde a_1|^2 + |\tilde a_2|^2)=0$. Using (\ref{eq:HE1}) and (\ref{eq:HE2}) to describe the derivatives, and, with reciprocal coupling already assumed, we find that 
\begin{equation} \label{eq:EnergyCons}
    g_{\rm s}\left|\tilde{a}_1\right|^2 = \gamma\left|\tilde{a}_2\right|^2.
\end{equation}
This result can be used to analytically find the expression for total stored energy $W=|\tilde a_1|^2(1+g_{\rm s}/\gamma)$. 

Another way to find the relationship between $\tilde a_1$ and $\tilde a_2$ is to directly solve for them from (\ref{eq:HEMatrix}) for a set of given parameters, and a specified steady-state pair. This method can determine the complex relation (phase and magnitude difference) between $\tilde a_1$ and $\tilde a_2$, but cannot determine the absolute phase or magnitude of either. It is important to note that the relationship in (\ref{eq:EnergyCons}) can be directly found from manipulating (\ref{eq:HEMatrix}) as shown in Appendix \ref{Appendix:EigandEnergy}, confirming the balance of energy in the steady state.

\subsection{Gain model}
\label{Sec:GainModel}
A key implication of determining the saturable gain is that there exists a steady-state energy, $|\tilde{a}_1|^2$, in the first oscillator such that
\begin{equation} \label{eq:steadyStG}
    g(|\tilde{a}_1|) = g_{\rm s}.
\end{equation}
The total energy in the system for a given steady-state oscillation frequency $\omega$ is thus determined through (\ref{eq:gs}), (\ref{eq:EnergyCons}), and (\ref{eq:steadyStG}).
To compute the system's energy, $W$, and analyze the steady state's {\em stability}, a gain model must be defined. Following previous works \cite{Mohseni2025_WPT,Zhou2016_PTSymmBreak,Wang2019_Chiral}, we use a saturable gain model consistent with laser theory as presented in \cite{Siegman1986_Lasers}. Here, it is convenient to define it as
\begin{equation} \label{eq:NonLGainModel}
    g(|a_1|) = \frac{g_0}{1+c|a_1|^2} - \gamma_\mathrm{i},
\end{equation}
with $g_0$ and $c$ being specific coefficients of a practical implementation of an active gain component, and $\gamma_\mathrm{i}$ represents the intrinsic losses of the first resonator. As the energy $|a_1|^2$ grows, the gain decreases until the steady-state value $\tilde{a}_1$, corresponding to the saturated gain $g_{\rm s}$, is reached. 

\begin{figure}[t]
\begin{centering}
\includegraphics[width=3.0in]{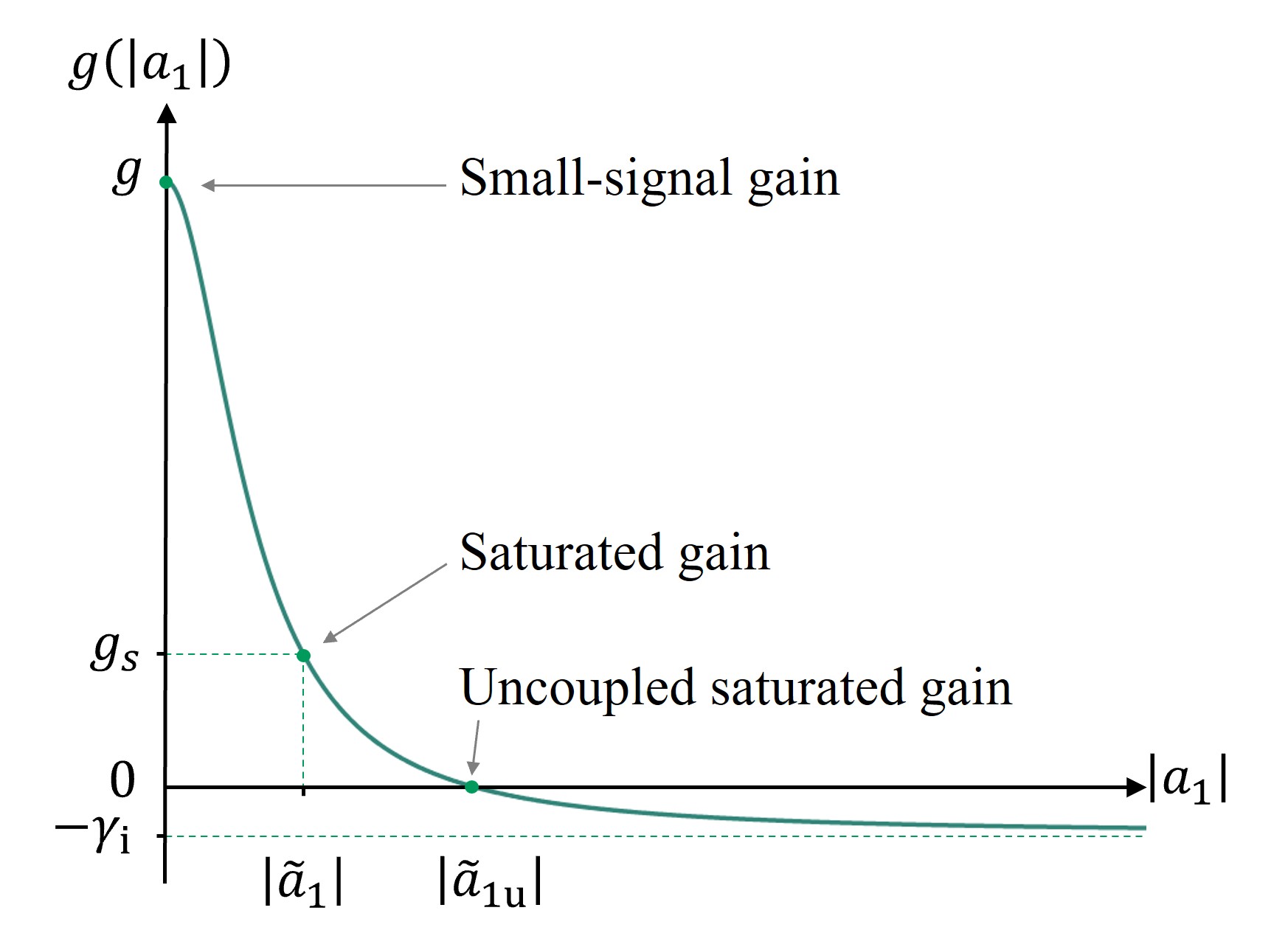}
\par\end{centering}
\caption{Plot of the nonlinear gain model in (\ref{eq:NonLGainModel}), including intrinsic losses $\gamma_{\mathrm{i}}$. The small-signal gain and coefficient $c$ are active gain component specific values, while the saturated gain $g_{\mathrm{s}}$ is a system-dependent value. Consequently, the steady-state amplitude $|\tilde{a}_1|$ depends on both the specific active device and the two-resonator system.} 
\label{fig:NonLGainModel}
\end{figure}

The gain versus state amplitude $|a_1|$ for this model is shown in Fig.~\ref{fig:NonLGainModel}, including realistic points. The small-signal gain is defined as $g(|0|)=g_0-\gamma_\mathrm{i}$; the saturated steady-state value, $\tilde{a}_{1}$ is found from (\ref{eq:steadyStG}), when the saturated gain, $g_{\rm s}$, is calculated from Sec.~\ref{sec:SaturatedGain}; the "uncoupled saturated gain" point $g(|\tilde{a}_{1\mathrm{u}}|)=0$ is what makes the first resonator, when uncoupled, steady.

Unlike previous works, the inclusion of $c$ in this model balances the dimensional analysis, as it has the units of inverse joules, and frees the values of $g_0$ and $\gamma_\mathrm{i}$ to take on any realistic values. This coefficient, $c$, determines how fast the steady regime is reached, and can be found from a point on the curve (most simply the uncoupled saturated gain point) if not directly available. It is important to note that the gain model should be independent from the definitions of $g_{\rm s}$ and $|\tilde{a}_1|$ as these are determined by the coupled system using the active component.

Utilizing (\ref{eq:steadyStG}) and (\ref{eq:NonLGainModel}), we obtain the steady-state amplitude $|\tilde{a}_1|$ for this gain model
\begin{equation} \label{eq:atilde}
    |\tilde{a}_1| = \sqrt{\frac{1}{c} \left(\frac{g_0}{g_{\rm s} + \gamma_\mathrm{i}} - 1\right)}.   
\end{equation}
This relation is valid only if $g_0 - \gamma_\mathrm{i} \ge g_{\rm s}$, which ensures that $g_{\rm s}$ lies within the active device’s gain range. 
In the weakly-coupled regime ($\kappa<\gamma$) that is one of the preferred ways to realize a highly sensitive oscillator (especially if $\kappa \approx \gamma$), the maximum saturated gain $g_{\rm s}=\gamma$ is found when $\omega_2=\omega_1$, as shown in Figs.~\ref{fig:ToySolGs}(b) and (c). Under this SS$\mathcal{PT}$ symmetry, $g_{\rm s}$ is maximized, and the signal amplitude $|\tilde{a}_1|$ assumes its smallest value.

Using (\ref{eq:atilde}), the total energy in the two resonators, $W$, for this specific gain model, is expressed in terms of $g_{\rm s}$ as
\begin{equation} \label{eq:Energy}
    W = \frac{1}{c}\left(\frac{g_0}{g_\mathrm{s} + \gamma_\mathrm{i}} -1\right)\left(1 + \frac{g_\mathrm{s}}{\gamma}\right).  
\end{equation}
Note that $W$ increases as $g_{\rm s}$ decreases and is maximized when $g_{\rm s}=0$, which occurs in the limit of zero coupling ($\kappa = 0$). 
As discussed in Sec.~\ref{sec:SaturatedGain}, under the weakly-coupled regime ($\kappa<\gamma$), the saturated gain assumes it maximum value  $g_{\rm s}=\gamma$ that corresponds to the minimum energy stored in the system as can be seen by plotting $W(g_{\rm s})$; Under this SS$\mathcal{PT}$-symmetry condition, the energy is equal to  $W_{\mathrm{SS}\mathcal{PT}} = \frac{2}{c}\left[g_0/(\gamma + \gamma_\mathrm{i}) -1\right]$. From this formula, we infer that though working with large $\gamma$ increases the dynamic range of high sensitivity, as seen in Fig.~\ref{fig:SensitivityRatio},  it decreases the signal amplitude.

\begin{figure}[t]
\begin{centering}
\includegraphics[width=3.5in]{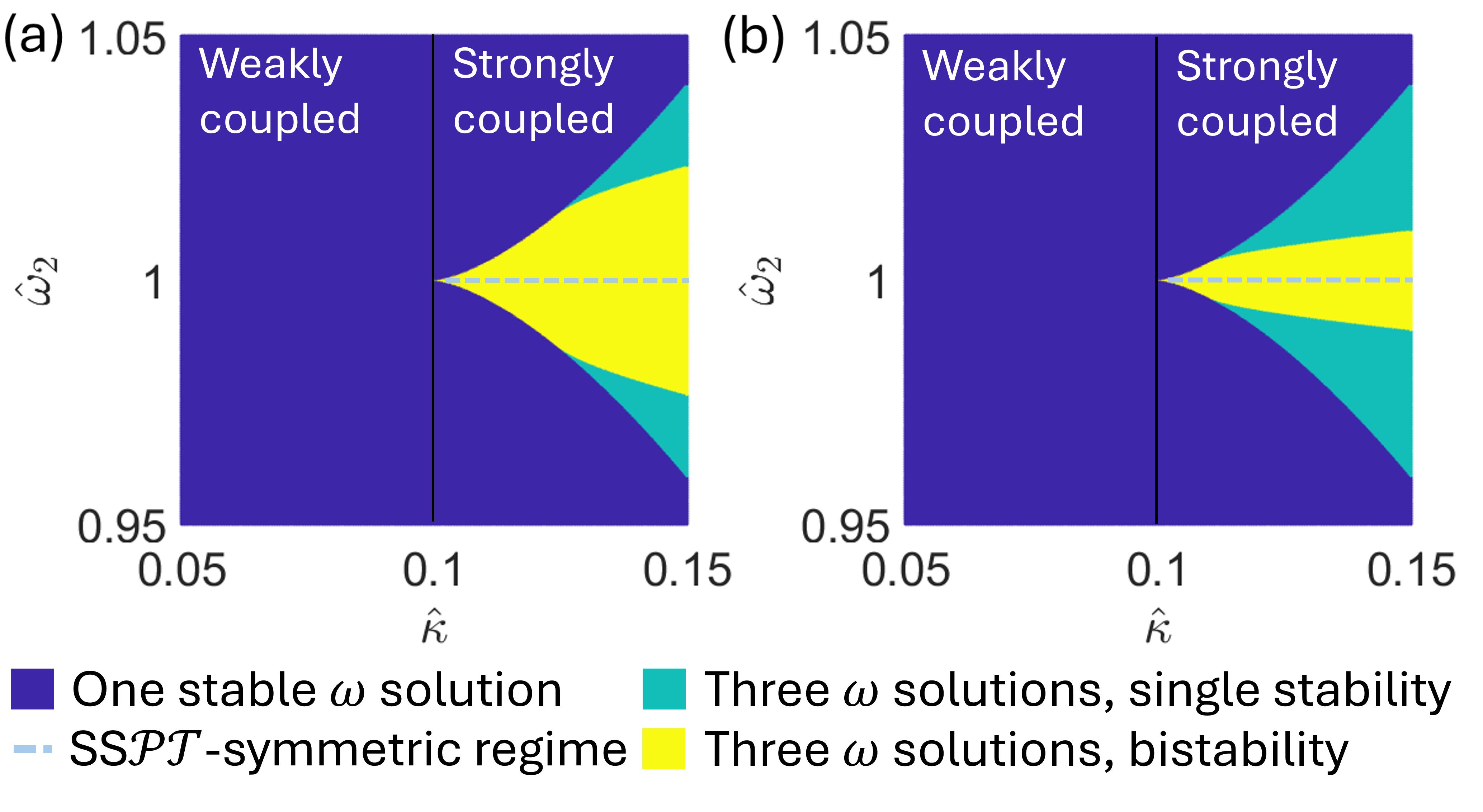}
\par\end{centering}
\caption{The solution regions and their stability corresponding to the $\omega$ and $g_{\rm s}$ values plotted in Fig.~\ref{fig:threeD} for a gain model with $c=1~\mathrm{J}^{-1}$, and $\hat{\gamma}_\mathrm{i} = 0.02$. The $\hat g_0$ values for the two plots are (a) $\hat g_0 = 0.2$, (b) $\hat g_0 = 0.15$. While the saturated steady-state oscillation  $\omega$ and gain $g_{\rm s}$ do not depend on the gain model, the shape of the stability regions does.
\label{fig:StabilityRegions}}
\end{figure}

\subsection{Stability analysis} \label{sec:Stability}

With a gain model specified, we analyze the stability of each steady-state pair $\omega$ and $g_{\rm s}$ by evaluating its associated Lyapunov exponents, following the approaches in \cite{Zhou2016_PTSymmBreak,Assawaworrarit2017_robust, Mohseni2025_WPT}, as detailed in the Appendix \ref{Appendix:Stability}. This analysis reveals that, for a given set of parameters, there exists either a single or two stable steady-state $\omega$ and $g_{\rm s}$ values. We refer to the region in parameter space supporting two stable steady-state $\omega$ and $g_{\rm s}$ as "bistable" \cite{Wang2019_Chiral,Dong2021_BistableWPT}. 

The single-stable and bistable regions in the parameter space are shown in Fig.~\ref{fig:StabilityRegions}, along with the regions of one or three steady-state frequencies mentioned in Sec.~\ref{sec:Steady-State Freq}, for two different sets of gain model values. The colors correspond to three different types of regimes, varying $\omega_2$ and $\kappa$: the blue area corresponds to the the existence of a single-stable SS oscillation frequency; the turquoise area corresponds to a single-stable SS oscillation frequency while the other two SS are unstable; and the yellow area corresponds to bistability where two SS frequencies are stable and the third one is unstable.

When multiple steady-state $\omega$ exist, the steady-state stable or bistable $\omega$ are always found to be connected to the lowest, or two lowest, $g_{\rm s}$ values thus making the middle steady-state $\omega$ always unstable (the $\omega$ closest to $\omega_1$, see Fig.~\ref{fig:threeD}, Fig~\ref{fig:ToySol}(e), and the largest $g_{\rm s}$ in Fig.~\ref{fig:ToySolGs}(d)). For all sets of parameters tested, at least a single-stable steady-state $\omega$ exists, except at the third-order SS-EPD that is seen as the separator regime between two regimes, one with one and the other with three SS oscillation frequencies, i.e., between the single-stable and the bistable regions (Fig.~\ref{fig:SymmetricSolutions}, Fig~\ref{fig:ToySol}(d), and Fig.~\ref{fig:ToySolGs}(c)).

As long as $g_0 - \gamma_i \ge g_{\rm s}$ (i.e., there is enough gain to support SS oscillations), the bistable region exists within the region of three SS frequencies. This is illustrated in the parameter-space plot in Fig.~\ref{fig:StabilityRegions}, where the bistable region (yellow area) is a subset of the region supporting three SS frequencies (yellow and turquoise areas), and both lie within the strongly coupled regime. This bistable region always includes the case with $\omega_2=\omega_1$, where the two stable SS frequencies, having $g_{\rm s}=\gamma$, form the "SS$\mathcal{PT}$-symmetric" regime (dashed line), and expands symmetrically across $\omega_2=\omega_1$. The size of the bistable region depends on the gain model values, and increases as $g_0$ increases, as seen from comparing Fig.~\ref{fig:StabilityRegions}(a) with Fig.~\ref{fig:StabilityRegions}(b). Corresponding results were found in \cite{Wang2019_Chiral}.

With this stability analysis, the SS oscillation frequencies $\omega$ shown in Fig.~\ref{fig:threeD} can be related to catastrophe theory \cite[pp.~70-74]{Strogatz2000_Chaos}. The third-order SS-EPD, the cusp in Fig.~\ref{fig:StabilityRegions}, exists exactly when the two bifurcation curves separating the steady-state regions meet tangentially \cite{Moncada2025sensorsOscill}. Moreover, the steady-state oscillation frequencies may undergo abrupt, discontinuous changes when a system's parameter is tuned outside the bistable region, leading to the formation of hysteresis loops \cite{Dong2021_BistableWPT, Moncada2026_Hysterisis}. The system also exhibits chiral behavior \cite{Wang2019_Chiral}, and both bistable frequencies can be accessed through slow, steady-state tuning of the parameters.

\subsection{Time domain analysis}
\label{TD analysis}
\begin{figure}[t]
\begin{centering}
\includegraphics[width=3.4in]{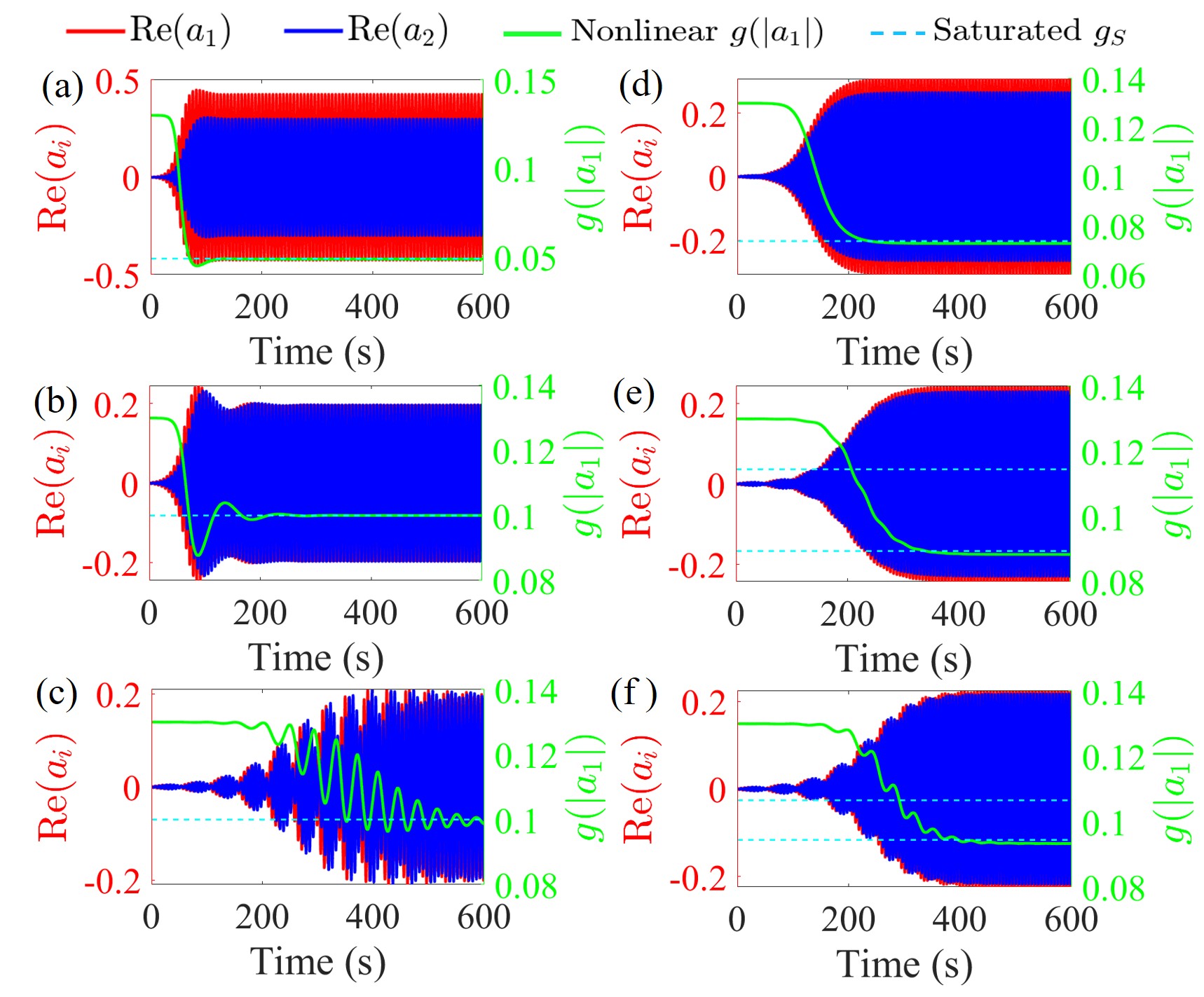}
\par\end{centering}
\caption{Time-domain simulations starting from the initial state $a_1 = a_2 = 0.001~\mathrm{J}^{1/2}$ at $t=0$ with a timestep of $10^{-3}\ \mathrm{s}$. For each time step, the real part of the state values are plotted (left ordinate axis), along with the nonlinear gain value in comparison with the expected $g_{\rm s}$ values (dashed green). For all plots, $\hat \gamma=0.1$, $\hat g_0=0.15$, $\hat \gamma_\mathrm{i}=0.02$, and $c = 6.5~\mathrm{J}^{-1}$, where the $\hat{\ }$ denotes a normalization to $\omega_1$. (a)-(c) Symmetric cases with $\hat\omega_2=1$, corresponding to points on the abscissa of Fig.~\ref{fig:SymmetricSolutions}, and: (a) $\hat \kappa = 0.07$, weakly-coupled case; (b) $\hat \kappa = 0.1=\hat\gamma$, third-order SS-EPD case with SS$\mathcal{PT}$ symmetry; (c) $\hat \kappa = 0.13$, strongly-coupled case. (c)-(f)  Strongly coupled cases with $\hat \kappa = 0.13 >\hat \gamma$, corresponding to points on the abscissas of Fig.~\ref{fig:ToySol}(e) and Fig.~\ref{fig:ToySolGs}(d), and: (c) $\hat \omega_2 = 1$ (a point on the abscissa of all previous plots), SS$\mathcal{PT}$-symmetric case because $g_{\rm s}=\gamma$; (d) $\hat \omega_2 = 0.97$; (e) $\hat \omega_2 = 0.99$; (f) $\hat \omega_2 = 0.995$.}
\label{fig:TDSimulations}
\end{figure}

To study the system's evolution and validate our analysis, we perform time-domain simulations of (\ref{eq:HE1}) and (\ref{eq:HE2}) using the gain model specified in (\ref{eq:NonLGainModel}). We employ a simple finite-difference scheme, the Forward Euler method, which requires small time steps because the governing equations are stiff \cite{LeVerque2007_FDMethods}. The six plots included in Fig.~\ref{fig:TDSimulations} show several representative simulations and validate the predicted steady-state frequencies, saturated gain, and the energy balance between the oscillators. In these simulations, without loss of generality, we choose $\omega_1=1~\mathrm{s}^{-1}$, necessitating the simulation length of $600~\mathrm{s}$. The hat $\hat{}$ of all the other parameters indicates normalization with respect to $\omega_1$. In the bistable region, the simulations consistently show that the oscillation frequency converges to the steady-state oscillation frequency $\omega$ associated with the lowest of the two stable saturated gains, $g_{\rm s}$. An example of multiple real saturated gain values is shown in Fig. \ref{fig:ToySolGs}(d).

The sets of parameters for each simulation in Fig.~\ref{fig:TDSimulations} were selected to probe the system’s behavior across the different solution regimes. The parameters for simulations (a)-(c) have $\omega_2=\omega_1$, and correspond to $\kappa$ points along the abscissa of Fig.~\ref{fig:SymmetricSolutions}, spanning from weak ($\hat \kappa=0.07< \hat{\gamma}$) to strong ($\hat \kappa=0.13 >\hat{\gamma}$) coupled regimes, including (c) the special regime with $\kappa=\gamma$ where three $\omega$ solutions form a third-order SS-EPD, which is one case of SS$\mathcal{PT}$-symmetric regime. The simulations (c)-(f), all within the strongly coupled regime ($\hat{\kappa}=0.13$), correspond to $\omega_2$ points along the abscissa of Fig.~\ref{fig:ToySol}(e) where bistability is observed. For all simulations, the transient duration varies with parameter choice, and this duration is directly linked to the stability of the steady-state frequencies through their Lyapunov coefficients, as defined in Appendix \ref{Appendix:Stability}.

Simulations (e), (f), and (c) fall within the bistable region, as indicated by the presence of multiple $g_{\rm s}$ values. In simulations (e) and (f), the system converges from its initial conditions to the lower of the two $g_{\rm s}$ values and to its corresponding oscillation frequency. Simulation (c) is distinct in that $\omega_2=\omega_1$ and it is in the "steady-state $\mathcal{PT}$-symmetric" regime, because the saturated gain $g_{\rm s}=\gamma$ and it is the same for both stable SS frequencies, hence it is degenerate; both stable (nondegenerate) SS frequencies are equivalently stable because they require the same (degenerate) saturated gain $g_{\rm s}=\gamma$. 
For the parameters in (c), there is also a SS oscillation frequency $\omega=\omega_1=\omega_2$ associated with the largest value of saturated gain in Fig.~\ref{fig:ToySolGs}(d); however, this solution is unstable.
For these simulations, the SS$\mathcal{PT}$-symmetric regime, besides the third-order degenerate condition, is generally the least stable regime in parameter space (requires the longest initial transient as does simulation (c)), as all other points in parameter space have at least one more stable steady-state oscillation frequency with $g_{\rm s}<\gamma$, see Sec. \ref{sec:SaturatedGain}.

We conclude by observing that the smallest signal amplitudes in Fig.~\ref{fig:TDSimulations} occur when $g_{\rm s}$ is maximized (i.e., when $g_{\rm s}=\gamma$), in agreement with what has been demonstrated at the end of Sec.~\ref{Sec:GainModel}.

\section{Inductively coupled circuits}  \label{sec:Inductive}

The nonlinear steady-state CMT analysis is applied to a pair of inductively coupled RLC circuits, as shown in Fig.~\ref{fig:IndCirFig}(a), and the results are compared with the nonlinear circuit analysis in \cite{Nikzamir2022_HighlySensitive}.

\subsection{General coupled oscillator approximations}

In connecting the general CMT equations for two coupled oscillators, (\ref{eq:HE1}) and (\ref{eq:HE2}), to a realistic circuit defined by circuit equations \cite{Nikzamir2022_HighlySensitive}, a difficulty arises in defining the CMT parameters $\omega_1$, $\omega_2$, $\kappa$, and $\gamma$ in terms of the circuit components $G_1,G_2,C_1,C_2,L,M,$ and the mutual inductive coupling $k$.  Establishing this connection is essential for predicting the behavior of the coupled circuits, as these parameters are not linearly related to the physical components. For uncoupled RLC circuits, the parameters can be defined exactly, as shown in Appendix \ref{Appendix:ExactCMT}. However, when coupling is introduced, only approximate expressions can be obtained.

For this reason, we propose the following approximations, as found in Appendix \ref{Appendix:IndCoupledCirc}. For the inductively coupled circuits in Fig.~\ref{fig:IndCirFig}(a), the CMT approximations are
\begin{equation} \label{eq:InductiveCoupledFirstApprox}
    \gamma \approx \frac{G_2}{2C_2},
\end{equation}
\begin{equation}
    \omega_{1} \approx \sqrt{\frac{1}{L C_1(1-k^2)} - \left(\frac{G_1}{2C_1}\right)^2},
\end{equation}
\begin{equation} \label{eq:InductiveCoupledLastApprox}
    \omega_2 \approx \sqrt{\frac{1}{L C_2(1-k^2)} - \gamma^2}.
\end{equation}
These approximations are only valid under the condition $k \ll 1$, which is an inherent limitation of the coupled-mode formalism applied to RLC circuits \cite{Haus1984_WavesFields}.

The coupling parameter $\kappa$ is found by imposing a third-order SS-EPD in the system: from the CMT, we know that to have a third-order SS-EPD, we must have $\kappa=\gamma$. Since the circuit in Fig.~\ref{fig:IndCirFig}(a) with parameters given in the following has a third-order SS-EPD as shown in \cite{Nikzamir2022_HighlySensitive} numerically, we choose  $\kappa =\gamma$.

\subsection{Results}
To obtain the results from the nonlinear SS-CMT analysis for this circuit, we apply the approximations from (\ref{eq:InductiveCoupledFirstApprox})–(\ref{eq:InductiveCoupledLastApprox}), using the same circuit parameters as \cite{Nikzamir2022_HighlySensitive}, to find $\kappa, \gamma, \omega_1,$ and $\omega_2$, and input these values into $p(\omega)=0$ to calculate the expected steady-state frequencies, where $p(\omega)$ is found in (\ref{eq:cubic}). The fixed circuit values are $G_1=G_2 = 20.52~\mathrm{mS}$, $L=0.1\ \mathrm{\mu H}$, $C_1 = 1~\mathrm{nF}$, and $k=0.2$ with $C_2$ treated as a variable parameter around the degenerate condition $C_2=C_1$ that provides a $\mathcal{PT}$-symmetric system (i.e., with $g_{\rm s}=\gamma$). From these, the nonlinear approximations for the inductively coupled circuit at the third-order degenerate condition are $g_{\rm s}=\gamma=\kappa = 1.026 \times10^7~\mathrm{s}^{-1}$ and $\omega=\omega_1=\omega_2=1.015\times10^8~\mathrm{s}^{-1}$. A valid nonlinear gain model is assumed.

\begin{figure}[t]
\begin{centering}
\includegraphics[width=3in]{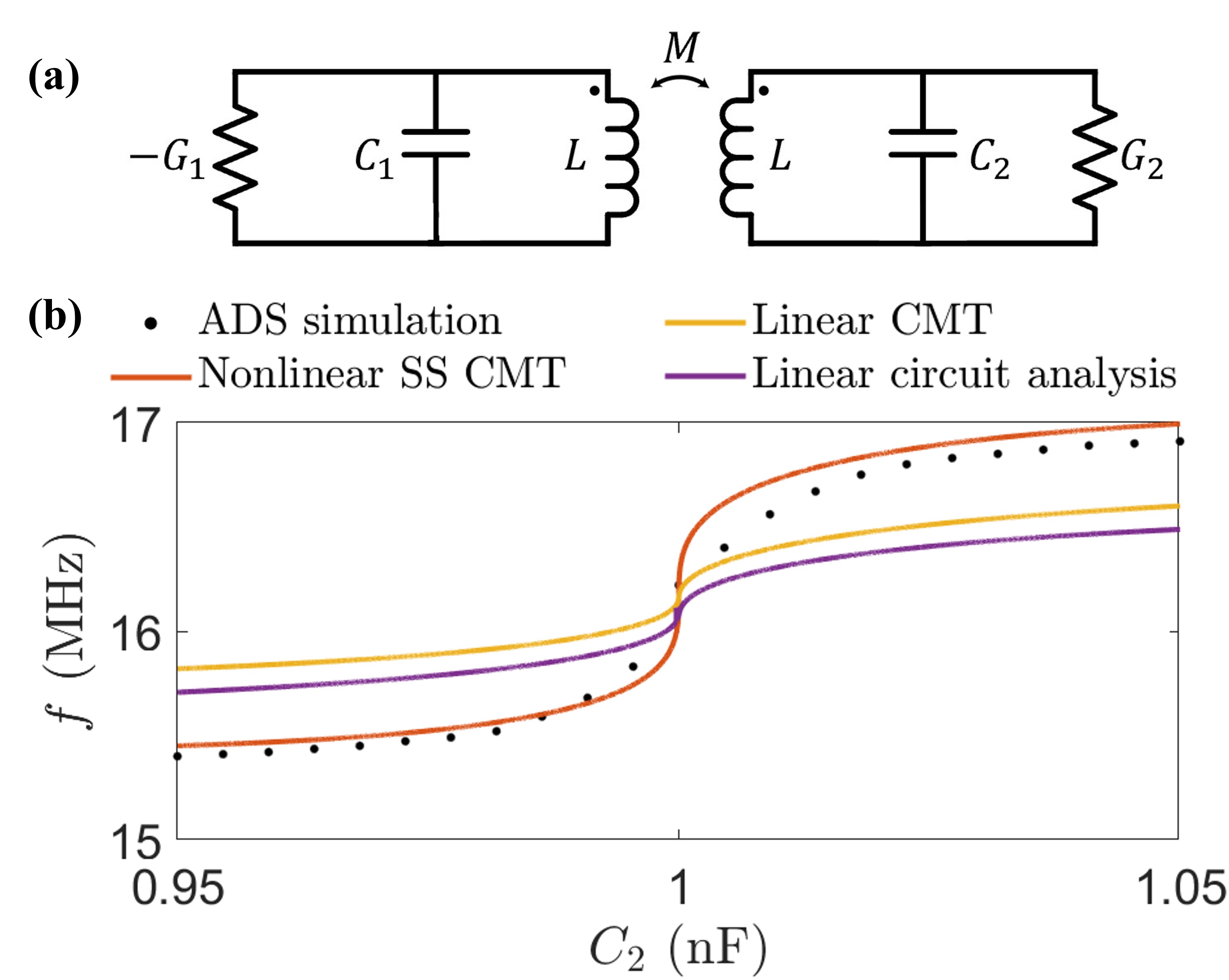}
\par\end{centering}
\caption{(a) Circuit diagram for two inductively coupled resonators with nonlinear gain. (b) Steady-state oscillation frequency versus perturbation of capacitor $C_2$ using two nonlinear calculations: time-domain ADS circuit simulator using the nonlinear $i-v$ curve; nonlinear steady-state (SS)-CMT theory of this paper using the approximations in (\ref{eq:InductiveCoupledFirstApprox}) - (\ref{eq:InductiveCoupledLastApprox}). These nonlinear calculations are in agreement. The results are compared with the $\mathrm{Re}(f)$ from the linear analysis in \cite[Fig. 4]{Nikzamir2022_HighlySensitive}
and from the linear CMT that provides the resonant frequency of the circuit.
}
\label{fig:IndCirFig}
\end{figure}

In Fig.~\ref{fig:IndCirFig}(b), we compare these results against other analyses and simulations of the same circuit: the linear circuit analysis detailed in \cite{Nikzamir2022_HighlySensitive}, a linear version of the CMT analysis, and time-domain simulations of the nonlinear circuit using the commercial software Keysight Advanced Design System (ADS). 

For the "Linear circuit analysis", an ideal small-signal linear gain $G_1 = 20.52~\mathrm{mS}$ is assumed. For the "Linear CMT" analysis, we make a similar assumption and use the ideal small-signal linear gain related to the small-signal circuit gain $G_1 = 20.52~\mathrm{mS}$; therefore, $g$ is constant and equal to $g \approx \frac{G_1}{2C_1}$. For the two linear analyses, we find the eigenfrequencies from (\ref{eq:HE1}) and (\ref{eq:HE2}). The linear analyses provide complex-valued frequencies, and only the real part is plotted in Fig.~\ref{fig:IndCirFig}(b).
For the time-domain ADS circuit simulation, the nonlinearity of the gain is described using the cubic current–voltage relation $i = -G_1 v + \alpha v^3$ for the negative conductance element, assuming the same values as in \cite{Nikzamir2022_HighlySensitive}, $G_1=1.001 G_2$ and $\alpha = 6.84~\mathrm{mS/V}^2$.

The comparisons show that the nonlinear SS-CMT analysis of this paper, though not exact when applied to such a circuit, provides results very close to the accurate time-domain ADS simulations of the circuit with a prescribed nonlinear $i-v$ curve. The SS-CMT model captures the behavior of the circuit, correctly predicting the high cubic-root sensitivity observed in the simulated result. 

The limitations of the CMT approximation of the circuit, using the parameters (\ref{eq:InductiveCoupledFirstApprox})-(\ref{eq:InductiveCoupledLastApprox}), are evident from comparing the two linear circuit analyses. The "Linear CMT" analysis is able to correctly predict the behavior of a linear circuit ("Linear circuit analysis") with only a slight frequency shift.

\section{Capacitively coupled circuits} \label{sec:Capacitive}
Next, we apply the nonlinear SS-CMT model described in this paper to a pair of capacitively coupled RLC circuits as shown in Fig.~\ref{fig:CapCirFig}(a). A complementary analysis of this nonlinear circuit, performed directly using nonlinear circuit equations, is available in \cite{Nikzamir2022_HighlySensitive, Moncada2025sensorsOscill}. While the nonlinear SS-CMT model is less precise, it lends significant intuition in understanding the phenomena observed in this paper.

\subsection{General nonlinear Hamiltonian approximation}
As with the inductively coupled RLC circuits, the main challenge with using the CMT approach lies in defining $\omega_1$, $\omega_2$, $\kappa$, and $\gamma$ in terms of the circuit components $G_1,G_2,C_1,C_2,L,$ and $C_c$, shown in Fig.~\ref{fig:CapCirFig}(a). Once again, we propose approximations found through the same method, as shown in Appendix \ref{Appendix:CapCoupledCirc}, and then impose $\kappa=\gamma$ to have the third-order SS-EPD:
\begin{equation} \label{eq:CapCoupledFirstApprox}
    \gamma \approx \frac{G_2}{2 C_2}\frac{B_1}{A },
\end{equation}
\begin{equation}
    \omega_{1} \approx \sqrt{\frac{B_2}{A L C_1} - \left(\frac{G_1}{2 C_1}\frac{B_1}{A }\right)^2},
\end{equation}
\begin{equation} \label{eq:CapCoupledLastApprox}
    \omega_2 \approx \sqrt{\frac{B_1}{A L C_2} - \gamma^2},
\end{equation}
where $A=1+C_c/C_1 + C_c/C_2$, $B_1 = 1 + C_c/C_1$, and $B_2 = 1 + C_c/C_2$. These approximations are valid when $C_c/C_1 \ll 1$ and $C_c/C_2 \ll 1$, i.e., for small capacitive coupling.

\subsection{Results}

The values of the parameters of the capacitively coupled circuit used in \cite{Nikzamir2022_HighlySensitive}, with ($C_c/C_1 = 1$), while producing the third-order SS-EPD, lie well outside of the valid region of the nonlinear SS-CMT analysis ($C_c/C_1 \ll 1$), causing this analysis to be a poor approximation of the actual circuit behavior. Therefore, to better validate the nonlinear SS-CMT analysis applied to the capacitively coupled circuit, we study the same circuit topology with a smaller capacitive coupling.

The chosen circuit values are $G_1=G_2=0.079~\mathrm{mS}$, $C_1=1.5~\mathrm{nF}$, $C_c=0.1~\mathrm{nF}$, and $L=10~\mathrm{\mu H}$ with $C_2$ again treated as a variable parameter around $C_2=C_1$ to form a third-order SS-EPD with a SS$\mathcal{PT}$-symmetry (i.e., with $g_{\rm s}=\gamma$). For the nonlinear time-domain circuit simulation using the commercial Key ADS package, we assume $G_1=1.001 G_2$ and $\alpha = 6.84~\mathrm{mS/V}^2$. The calculated values for the nonlinear analysis at the third order degenerate solution are $g_{\rm s}=\gamma=\kappa = 2.48*10^5~\mathrm{s}^{-1}$ and $\omega=\omega_1=\omega_2= 7.92*10^6~\mathrm{s}^{-1}$.

\begin{figure}[t]
\begin{centering}
\includegraphics[width=3in]{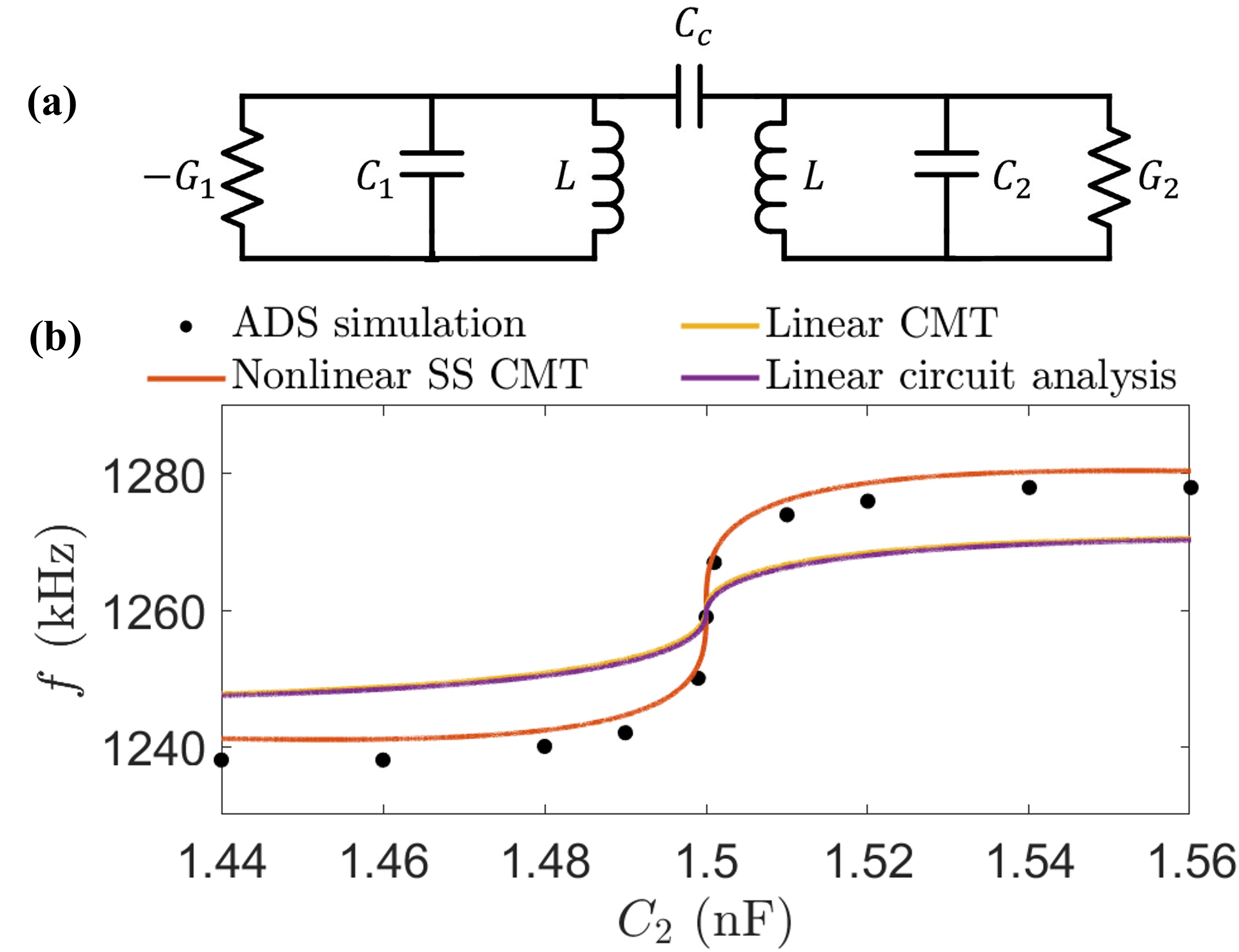}
\par\end{centering}
\caption{As in Fig~\ref{fig:IndCirFig}, except that now (a) the two RLC circuits are coupled through a capacitor, and (b) the SS-CMT theory uses the approximations in (\ref{eq:CapCoupledFirstApprox}) - (\ref{eq:CapCoupledLastApprox}). The nonlinear SS CMT results show good agreement with those from the TD ADS nonlinear circuit simulations.
}
\label{fig:CapCirFig}
\end{figure}

The comparison in Fig.~\ref{fig:CapCirFig}(b) shows good agreement between the oscillation frequencies calculated using ADS simulations and the general nonlinear SS-CMT analysis, including the prediction of the third-order SS-EPD. In this circuit, both the coupling and losses are much smaller than in the previously considered inductively coupled circuit example. This greatly increases the accuracy of the approximate CMT method, as shown by how closely the "Linear CMT" model matches the "Linear circuit analysis" results. It is also important to note that low losses (small $\gamma$) reduce the range of frequencies of high sensitivity, consistent with the trends observed in Fig.~\ref{fig:SensitivityRatio}.

\section{Conclusion}
We have analyzed the steady-state (SS) behavior of the nonlinear system of coupled oscillators depicted in Fig. \ref{fig:CoupledOsc}, using coupled-mode theory (CMT) and focusing on its oscillation frequencies and saturated gain values. This system exhibits SS-EPDs of order two and three, which separate the regions in the parameter space of either single or multiple SS oscillation frequencies and saturated gain pairs. In particular, a unique third-order SS-EPD occurs in this system at $\omega=\omega_1=\omega_2$ and $g_{\rm s}=\gamma = \kappa$, at which the SS oscillation frequency has a square-root sensitivity to perturbations in $\kappa$ and $\gamma$ and a cubic-root sensitivity to perturbations in $\omega_1$ and $\omega_2$. We have also shown that the region of very high sensitivity with respect to $\omega_2$ (i.e., the dynamic range) increases when increasing losses $\gamma$; however, when increasing losses and working near the third-order SS-EPD, the signal strength decreases, as the saturated gain reaches its maximum. 

A key aspect of understanding the system’s behavior is the stability of the SS oscillation frequencies and the corresponding saturated gain values. To analyze this, we employed a specific saturable-gain model, which also allowed us to derive the energy characteristics of the system. Using this model, we investigated the stability and identified the associated regions: monostability and bistability. We find that both the strength of the stability and extent of the bistable region are directly linked to the saturated gain values and its symmetry in parameter space. 

Leveraging these insights, we applied this nonlinear SS-CMT analysis to coupled RLC circuits. Although approximate due to inherent limitations of CMT, the analysis shows good agreement with time-domain nonlinear circuit simulations and successfully captures the increased sensitivity of the nonlinear system to perturbations in its resonant frequencies.

In conclusion, we find that one of the most effective operating regimes for achieving high sensitivity to small perturbations of either  $\omega_1$ or $\omega_2$ is the weakly-coupled regime with $\kappa \approx \gamma$, where the sensitivity is approximately linear, e.g., $\Delta \omega \approx \alpha_1 \Delta \omega_2$. When $\kappa \approx \gamma$,  the coefficient $\alpha_1$ becomes very large and diverges as $(\gamma-\kappa) \rightarrow 0$, i.e., when converging to the third-order SS-EPD. The advantage of working close to, but not exactly at, the SS-EPD comes from having linear approximately sensitivity and, in particular, from avoiding the difficulty of operating precisely at the third-order SS-EPD. Indeed, even a slight increase in the coupling $\kappa$ drives the system into the strongly-coupled regime, where enhanced sensitivity occurs in hysteresis loops due to bistability. The hysteresis loops can be beneficial for other types of sensing not discussed here. 



\appendix

\section{Real $\omega$ solutions to (\ref{eq:OmegaSol})} \label{Appendix:A}  

We show an alternative way to find the real-valued oscillation frequencies $\omega$ by forcing the imaginary part of $\omega$ to be zero in (\ref{eq:OmegaSol}). 
Purely real solutions to (\ref{eq:OmegaSol}) only exist when 
\begin{equation}\label{eq:imag sqrt}
    \gamma-g_{\rm s}= \mp \mathrm{Im}\left[\sqrt{4\kappa^2 + \left[(\omega_1-\omega_2)-   j(\gamma + g_{\rm s})\right]^2}\right].
\end{equation}
The square root is split into its real and imaginary parts
\begin{equation}
    \sqrt{u + iv} = \sqrt{\frac{u+ \sqrt{u^2 + v^2}}{2}} + i\text{sgn}(v) \sqrt{\frac{-u+ \sqrt{u^2 + v^2}}{2}},
\end{equation}
with $u=4\kappa^2+(\omega_1-\omega_2)^2-(\gamma+g_{\rm s})^2$ and $v=-2(\omega_1-\omega_2)(\gamma+g_{\rm s})$. We simplify (\ref{eq:imag sqrt}) as
\[    \gamma-g_{\rm s}= \mp\text{sgn}(v) \sqrt{\frac{-u+ \sqrt{u^2 + v^2}}{2}},\]
\[    (\gamma-g_{\rm s})^2 = \frac{-u+ \sqrt{u^2 + v^2}}{2},\]
\begin{equation} \label{eq:Appendixeq3}
    4(\gamma-g_{\rm s})^4+4u(\gamma-g_{\rm s})^2- v^2 = 0.
\end{equation}    

Here, if $v = 0$, which is possible only when $\omega_1 = \omega_2$ or when $g_{\rm s}=\gamma=0$, (\ref{eq:Appendixeq3}) simplifies to
\begin{equation} \label{eq:ReCondition}
    (\gamma-g_{\rm s})(4\kappa^2-4\gamma g_{\rm s} + (\omega_1-\omega_2)^2) = 0.
\end{equation}
From (\ref{eq:ReCondition}) and with $v = 0$, we find that the oscillation frequency is only real if $g_{\rm s}=\gamma$ or if $\omega_1 = \omega_2$ and $g_{\rm s} = \kappa^2/\gamma$. These conditions are equivalent to those found in Sec.~\ref{sec:SymmetricProblem}, where we find the associated oscillation frequency values.

If $v \neq 0$, (\ref{eq:Appendixeq3}) expands out to be
\begin{equation} \label{eq:Appendixeq5}
    \gamma g_{\rm s}^3 - 2(\gamma^2+2\kappa^2)g_{\rm s}^2 + (\gamma_s^2+2\kappa^2+\gamma(\omega_1-\omega_2)^2)\gamma g_{\rm s} - \gamma_s^2\kappa^2 = 0.
\end{equation}
This equation is equivalent to the saturated gain equation (\ref{eq:gsCubic}) in Sec.~\ref{sec:SaturatedGain}. 

\begin{figure*}
\includegraphics[width=\textwidth]{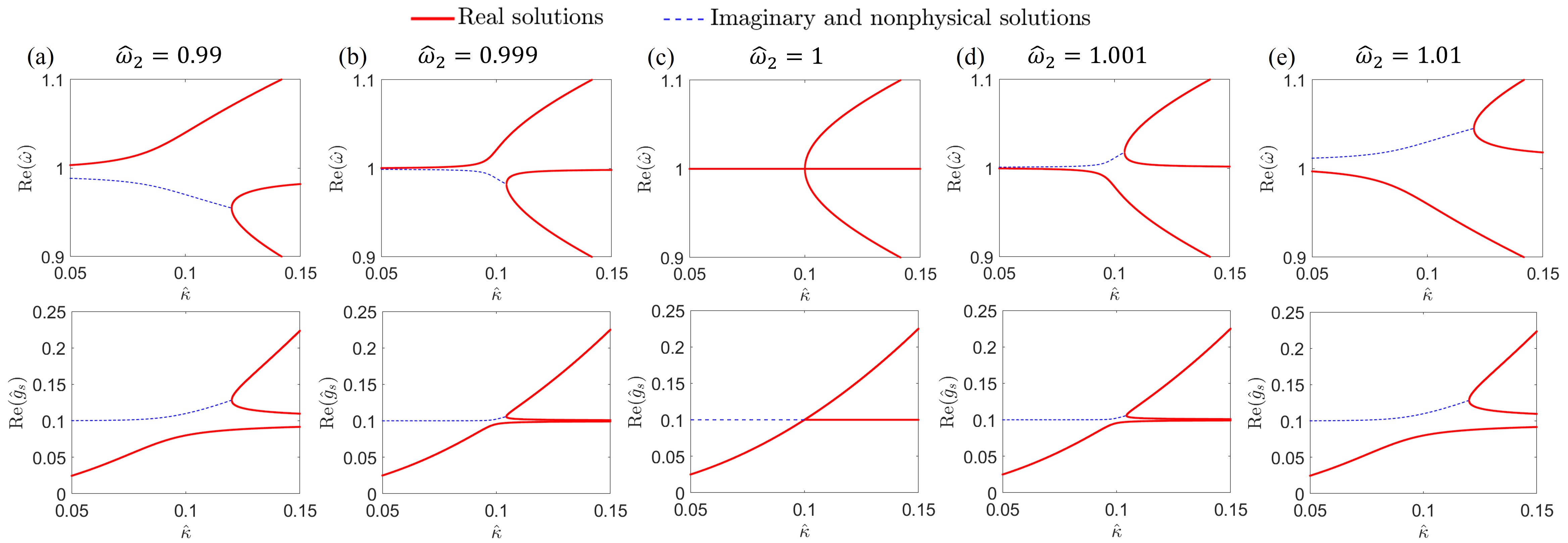}
\caption{Steady-state purely-real oscillation frequencies and their associated purely real saturable gain values (in red), plotted around the point $\omega_1=\omega_2$ and $\kappa=\gamma$ varying $\hat{\kappa}$ (the complex-valued branches, blue dots, are shown for a better understanding of the solutions). For all plots $\hat \gamma = 0.1$ and their $\hat \omega_2$ values are listed on top, with the $\hat{\ }$ denoting a normalization to $\omega_1$.
\label{fig:AppendAnalysis}}
\end{figure*}

\section{The third order degeneracy conditions of $p(\omega)=0$} \label{Appendix:3rd order degeneracy}

In the symmetric system, where $\omega_{\mathrm{i}}=\omega_1=\omega_2$ with $\omega_{\mathrm{i}}$ denoting a degenerate solution, we know that the special third-order SS-EPD occurs when $\kappa=\gamma$ as shown in Sec.~\ref{sec:SymmetricProblem}. To verify the order of this degeneracy, we study the cubic equation $p(\omega)=0$.
A third order degeneracy at $\omega=\omega_\mathrm{i}$ occurs when $p(\omega)=0$ when
\begin{equation}
p(\omega) = (\omega - \omega_{\mathrm{i}})^3.
\end{equation}
This further expands to
\begin{equation} \label{eq:3orderDeg}
p(\omega) = \omega^3 - 3\omega_{\mathrm{i}}\omega^2 + 3\omega_{\mathrm{i}}^2\omega - \omega_{\mathrm{i}}^3.
\end{equation}
We now find the conditions under which this occurs. Setting $b_2, b_1$ and $b_0$ equal to the coefficients in (\ref{eq:3orderDeg}) creates the following system of equations:
\begin{equation} \label{eq:EDP31}
\omega_{\mathrm{i}} = \frac{1}{3}\omega_1 + \frac{2}{3}\omega_2,
\end{equation} 
\begin{equation} \label{eq:EDP32}
\omega_2^2 + 2\omega_1\omega_2 + \gamma^2 - \kappa^2 - 3\omega_{\mathrm{i}}^2 = 0,
\end{equation}
\begin{equation} \label{eq:EDP33}
-\omega_1\omega_2^2 - \gamma^2\omega_1 + \kappa^2\omega_2 + \omega_{\mathrm{i}}^3 = 0.
\end{equation}
Inserting (\ref{eq:EDP31}) into (\ref{eq:ImCharacteristic}), we find that $(\omega_1-\omega_2)(2\gamma+g_{\rm s})=0$. This simple derivation yields the necessary condition for a third-order degeneracy to exist: $\omega_1=\omega_2$. The other condition, $g_{\rm s}=-2\gamma$, is not physical. Using $\omega_1=\omega_2$ in (\ref{eq:EDP31})-(\ref{eq:EDP33}), one finds that $\gamma=\kappa$, and thus the only third order degenerate condition for the system of coupled oscillators is the one with steady-state oscillation frequency $\omega = \omega_1=\omega_2$ and $g_{\rm s}=\gamma=\kappa$.

\section{Sensitivity derivation}
\label{Appendix:Sensitivity}  

\subsection{Cubic root sensitivity of $\omega(\omega_2)$ when $\gamma = \kappa$} \label{Appendix:w2(w)Sensitivity}

Instead of directly studying the function $\omega(\omega_2)$ and its sensitivity $d\omega/d \omega_2$, it is convenient to consider the inverse function $\omega_2(\omega)$, that satisfies $p(\omega,\omega_2(\omega))=0$ and look at its derivative, $d\omega_2(\omega)/d\omega$. Applying the implicit function theorem to this case, we find
\begin{equation} \label{eq:InverseImplDer}
\frac{d \omega_2(\omega)}{d\omega} = \frac{p^{\prime}(\omega)}{2 \omega^2 -2(\omega_1+\omega_2)\omega +2 \omega_1 \omega_2 - \kappa^2}.
\end{equation}
where here $\omega_2$ is a function of $\omega$. In the neighborhood of $\omega=\omega_1$, occurring only when $\omega_1=\omega_2$, 
we look at the perturbation $\Delta \omega_2=\omega_2-\omega_1$ of $\omega_2$,  holding $\omega_{1}$ constant. 
When $\omega = \omega_1$, we have $d\omega_2/d \omega=(\kappa^2-\gamma^2)/\kappa^2$ verifying that this function is differentiable across this point even when $\gamma=\kappa$.  
As this is the case, we assume that in the neighborhood of $\omega=\omega_1$ we have that $\omega_2 = \omega_1+ \Delta\omega_2$, and we expand the function $\Delta \omega_2$ in Taylor series for small $\Delta\omega$, 
\begin{equation} 
\Delta \omega_2=\alpha_1 \Delta \omega +\alpha_2 (\Delta\omega)^2+\alpha_3 (\Delta\omega)^3+O((\Delta\omega)^4),
\end{equation}
We also expand the numerator and denominator of (\ref{eq:InverseImplDer}) in these terms: the numerator simplifies to $p^{\prime}(\omega) = 3\Delta\omega^2 -4\Delta\omega_2\Delta\omega+\Delta\omega_2^2+\gamma^2-\kappa^2$, and the denominator simplifies to $2\Delta\omega^2 -2\Delta\omega_2\Delta\omega-\kappa^2$.

Using this expansion and considering also that the left hand side of (\ref{eq:InverseImplDer}) is $d\omega_2/d\omega=  \alpha_1  +2 \alpha_2 (\Delta\omega)+3 \alpha_3 (\Delta\omega)^2+O((\Delta\omega)^3)$, we obtain that 
\begin{equation} 
\label{eq:sensitivity-Coeffs1}
\alpha_1 = \frac{\kappa^2-\gamma^2}{\kappa^2}, \quad \alpha_2=0, \quad \alpha_3 = -\frac{\gamma^4}{\kappa^6},
\end{equation}
which leads to the approximation in (\ref{eq:TaylorSeriesExp}). Assuming that $\gamma=\kappa$ forces $\alpha_1=0$ and we find that $p^{\prime}(\omega) = 3(\omega-\omega_1)^2 +O((\omega-\omega_1)^3) $ and $p(\omega) = (\omega-\omega_1)^3 +O((\omega-\omega_1)^4)$.

Therefore, in the neighborhood of the third-order EPD (i.e., $\omega \approx \omega_1$), when $\gamma=\kappa$, $\Delta \omega_2 \approx (-1/\kappa^2) (\Delta \omega)^3$ that, when inverted, leads to $\Delta \omega \approx - \kappa  ^{2/3}(\Delta \omega_2)^{1/3}$, which is (\ref{eq:SensitivityDegenr}).

\subsection{Linear, super-high sensitivity of $\omega(\omega_2)$  when  $\gamma \ne \kappa$} \label{Appendix:w(w2)Sensitivity}

In this case, when $\omega_2=\omega_1$, we have that  $p^{\prime}(\omega=\omega_1) = (\gamma^2-\kappa^2) \ne 0$. Therefore, the sensitivity function  $d \omega/ d \omega_2$ in (\ref{eq:Sol1disersion equationNL}) is also differentiable at $\omega_2=\omega_1$ (i.e., when $\Delta \omega_2=0$). 

It is convenient to define $\omega=\omega_1+\Delta \omega$, where  $\Delta \omega$ is a function of $\omega_2$ that vanishes when $\omega_2=\omega_1$, as seen from (\ref{eq:ReCharacteristicNOg}) and $p=0$. The numerator of (\ref{eq:Sol1disersion equationNL}) is rewritten as $2 (\Delta \omega)^2-2\Delta \omega \Delta \omega_2 - \kappa^2$, whereas the denominator is $p^{\prime}= 3 (\Delta \omega)^2  - 4 \Delta \omega \Delta\omega_2 +(\Delta \omega_2)^2 +(\gamma^2-\kappa^2)$. In these two polynomials we use the Taylor series expansion of $\Delta \omega$ for small $\Delta\omega_2$ is
\begin{equation} 
\Delta \omega=\alpha_1 \Delta \omega_2 +\alpha_2 (\Delta\omega_2)^2+\alpha_3 (\Delta\omega_2)^3+O((\Delta\omega_2)^4).
\end{equation}
The Taylor expansion of the left hand sides of (\ref{eq:Sol1disersion equationNL}), at $\omega_2=\omega_1 + \Delta \omega_2$, is
\begin{equation}
\frac{d\omega}{d\omega_2}\bigg|_{\omega_1+\Delta \omega_2}=  \alpha_1  +2 \alpha_2 (\Delta\omega_2)+3 \alpha_3 (\Delta\omega_2)^2+O((\Delta\omega_2)^3),
\end{equation}
where $\alpha_1=\frac{d\omega}{d\omega_2}\big|_{\omega_1}$. Using these three expansion series in the left and right polynomials in (\ref{eq:Sol1disersion equationNL}), and comparing the coefficients, leads to 
\begin{equation} 
\label{eq:sensitivity-Coeffs2}
\alpha_1 = \frac{-\kappa^2}{\gamma^2-\kappa^2}, \quad  \alpha_2=0, \quad \alpha_3= \frac{ \kappa^2 \gamma^4}{(\gamma^2-\kappa^2)^4}  .
\end{equation}

This demonstrates the first order expansion of the sensitivity in (\ref{eq:sensitivity-Normal}). Note that the coefficients diverge when $\gamma \rightarrow \kappa$, as expected.

\subsection{Linear $\omega(\kappa)$ sensitivity, when $\omega_1 \ne \omega_2$} \label{Appendix:w(k)Sensitivity}
In Sec.~\ref{sec:sqrt-Sensitivity} we have previously found the sensitivity of the oscillation frequency $\omega(\kappa)$ to small variation of $\kappa$ around $\kappa = \gamma$ is a square-root, when $\omega_1 = \omega_2$:  here we will prove that when $\omega_1 \ne \omega_2$, the sensitivity of $\omega(\kappa)$  is mainly linear to $\Delta\kappa$, but still heightened when $\omega_1 \approx \omega_2$.

As in the previous two sections, we apply the implicit function theorem 
\begin{equation} \label{eq:ImplDerKappa}
\frac{d\omega(\kappa)}{d\kappa} = \frac{2\kappa(\omega-\omega_2)}{p^{\prime}(\omega)}.
\end{equation}
We rewrite the numerator and denominator of (\ref{eq:ImplDerKappa}) using the change of variable $\kappa = \gamma + \Delta\kappa$, and we expand the function $\omega(\gamma + \Delta\kappa)$ in Taylor series for small $\Delta\kappa$,
\begin{equation} 
\omega(\gamma+\Delta \kappa)=\omega_0 + \alpha_1 \Delta \kappa + O((\Delta\kappa)^2).
\label{eq:Deltaomegataylor}
\end{equation}
Here, $\omega_0=\omega(\kappa=\gamma)$, i.e., when $\Delta \kappa=0$, is the solution of $p(\omega)=0$ when $\kappa=\gamma$. 
Using the expansion also in the left hand side of (\ref{eq:ImplDerKappa}),  we find that $\frac{d\omega}{d\kappa}\big|_{\gamma+\Delta \kappa} = \alpha_1 +O(\Delta\kappa)$. Comparing the coefficients of the left and right hand sides of (\ref{eq:ImplDerKappa}) we find that
\begin{equation}
\alpha_1 = \frac{-2\gamma}{2\omega_1 + \omega_2 - 3\omega_0}. \\  
\end{equation}
This verifies that when $\omega_1\neq\omega_2$, the sensitivity of $\omega(\kappa)$ to small perturbations of $\kappa$ around $\kappa = \gamma$ is approximately linear. 

It is possible to find an analytic solution for the oscillation frequency $\omega_0$ in this regime, where $\omega_2\ne\omega_1$ and $\kappa=\gamma$. We apply the Cardano's Formula to solve the cubic equation $p(\omega_0)=0$, as this case always falls in the region with only a single steady-state $\omega$. Thus we find
\begin{equation} 
\omega_0 = \frac{\omega_1+2\omega_2}{3} + \sqrt[3]{Q+V} +  \sqrt[3]{Q -V},
\end{equation}
where 
\begin{equation}
\begin{split}
    Q =\frac{(\omega_1-\omega_2)^3}{27}+\gamma^2\frac{(\omega_1-\omega_2)}{2},\\
    V = \gamma(\omega_1-\omega_2)\sqrt{\frac{(\omega_1-\omega_2)^2}{27}+\frac{\gamma^2}{4}}.
\end{split}
\end{equation}
Note that when $\omega_2=\omega_1$, one has $Q=0$, $V=0$, and therefore $\omega_0=\omega_2=\omega_1$, and the coefficient $\alpha_1$ diverges, as expected.

\subsection{Sensitivity of $\omega(\omega_2)$. Alternative method} \label{Appendix:Taylor}

When consider the polynomial $p(\omega)=0$ as a function of two variables, $\omega$ and $\omega_2$, and we rewrite it using the change of variables $\omega=\omega_1+\Delta\omega$ and $\omega_2=\omega_1+\Delta\omega_2$. Applying a multivariable Taylor series expansion, or, alternatively through algebraic manipulation, we find
\begin{equation} 
\label{eq:MultivariateTaylorSeries}
\Delta\omega^3 - 2\Delta\omega^2\Delta\omega_2 + \Delta\omega\Delta\omega_2^2 + (\gamma^2-\kappa^2)\Delta\omega + \kappa^2\Delta\omega_2=0.
\end{equation}
When operating near the frequency $\omega_1$ ($\omega\approx\omega_1$ and $\omega_2\approx\omega_1$) with $\kappa\neq\gamma$, the linear terms are dominant, leading to $(\kappa^2-\gamma^2)\Delta\omega \approx \kappa^2\Delta\omega_2$ that shows the linear sensitivity in (\ref{eq:sensitivity-Normal}).

Instead, when operating near the third-order SS-EPD, i.e., $\omega_2\approx \omega_1$ with $\kappa=\gamma$ (hence also $\omega \approx \omega_1$) and looking for variations in $\omega$ to small changes in $\omega_2$, the cubic term and the term $\kappa^2 \Delta\omega_2$ dominate, leading to $\Delta\omega^3 \approx -\kappa^2\Delta\omega_2$ that shows the cubic-root sensitivity in (\ref{eq:SensitivityDegenr}).

This analysis simply verifies the sensitivity analysis performed previously, showing that around the third order degenerate solution the sensitivity of $\omega$ to changes in $\omega_2$ is cube-root like, while otherwise it is mainly linear. However, an important note is that the linear term, $(\gamma^2-\kappa^2)\Delta\omega$, in (\ref{eq:MultivariateTaylorSeries}) becomes less dominant as $\kappa$ approaches $\gamma$. This causes the sensitivity of $\Delta\omega$ to $\Delta\omega_2$ to still be increased and approach cube-root sensitivity in the neighborhood of $\gamma=\kappa$.

\section{Eigenvector and energy conservation equivalence} \label{Appendix:EigandEnergy}

The results from analyzing the energy in (\ref{eq:EnergyCons}) can also be obtained directly from manipulating (\ref{eq:HEMatrix}). Here, the symbol * indicates complex conjugation, whereas a dagger $\dagger$ indicates the Hermitian adjoint of a vector, and ${\rm T}$ indicates the transpose operation. In order to work with the energies of both oscillators, we multiply both sides of (\ref{eq:HEMatrix}) by the hermitian adjoint of $\boldsymbol{\tilde a} = [\tilde a_1,\tilde a_2]^{\rm T}$:
\begin{equation}
\omega \boldsymbol{\tilde a}^\dagger \boldsymbol{\tilde a} =
\boldsymbol{\tilde a}^\dagger
\begin{bmatrix}
\omega_1 - j g_{\rm s} & -\kappa  \\
-\kappa & \omega_2 + j\gamma
\end{bmatrix}
\boldsymbol{\tilde a} .
\end{equation}
This simplifies to 
\begin{equation} \label{eq:EigVandEnergyCons}
\begin{split}
    \omega (|\tilde a_1|^2 + |\tilde a_2|^2) - \omega_1|\tilde a_1|^2 - \omega_2|\tilde a_2|^2 + \kappa(\tilde a_1^* \tilde a_2 + \tilde a_2^* \tilde a_1) \\
    + j(g_{\rm s} |\tilde a_1|^2  - \gamma |\tilde a_2|^2) = 0.
\end{split}
\end{equation}
Since $\omega, \omega_1,\omega_2, \gamma, \kappa,$ and $g_{\rm s}$ are all real in the steady-state regime, and as the product $\tilde a_1^* \tilde a_2 + \tilde a_2^* \tilde a_1$ is also real, (\ref{eq:EnergyCons}) is found from the imaginary part of (\ref{eq:EigVandEnergyCons}).

\section{Stability analysis} \label{Appendix:Stability}
%
\begin{figure*}
  \includegraphics[width=\textwidth]{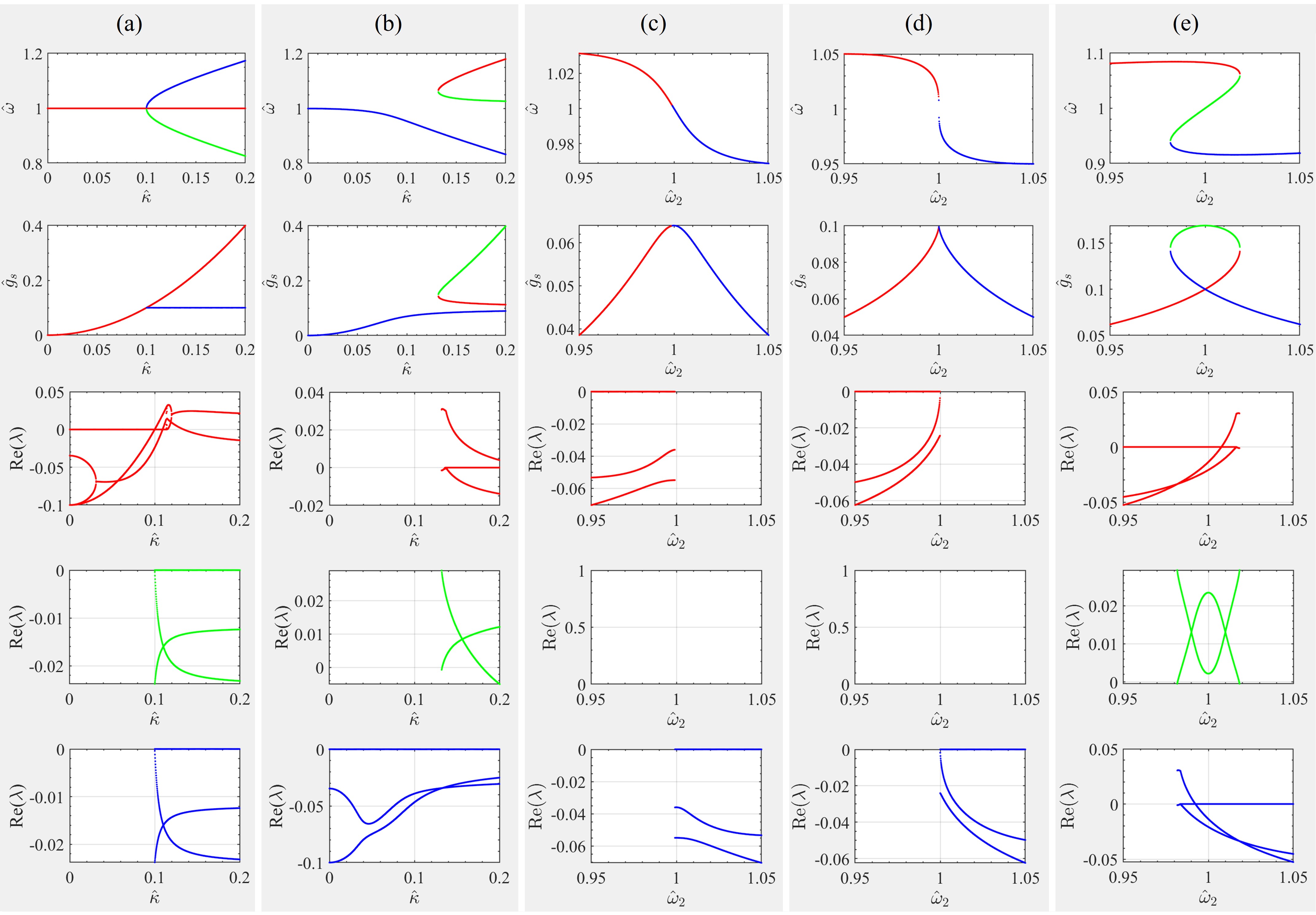}
  \caption{Each column shows the SS oscillation frequency $\omega$ and gain $g_{\rm s}$ and the real part of their four associated Lyapunov exponents plotted varying $\kappa$ or $\omega_2$. These cuts of the parameter space correspond those already shown in Fig.~\ref{fig:SymmetricSolutions}, Fig.~\ref{fig:ToySol}, and Fig.~\ref{fig:ToySolGs}, but with separate colors differentiating each unique solution. Each unique steady-state solution has an associated plot of its four Lyapunov exponents, leading to a total of three separate Lyapunov exponent plots. Though each plot has four Lyapunov exponents, their real parts may overlap causing them to be indistinguishable at points. In each plot $\hat{\gamma}=0.1$, $\hat{g}_0=0.15$, $\hat{\gamma}_\mathrm{i}=0.02$ and $c=1~\mathrm{J}^{-1}$, with the $\hat{\ }$ denoting a normalization to $\omega_1$. The parameters for each plots are as follows. Varying $\hat{\kappa}$: (a) $\hat{\omega}_2=1$; (b) $\hat{\omega}_2=1.02$. Varying $\hat{\omega}_2$: (c) $\hat{\kappa}=0.08$; (d) $\hat{\kappa}=0.1=\hat\gamma$; (e) $\hat{\kappa}=0.13$.
  \label{fig:StabilityAnalysis}}
\end{figure*}
The gain element in our system of coupled oscillators exhibits a saturable nonlinearity: as the mode amplitude $|a_1|$ increases, the gain saturates, leading to steady-state oscillations. Among the available resonant modes, the mode requiring the lowest gain will dominate, reaching its steady state and saturating the gain, which prevents other modes from accessing the necessary gain to achieve other steady-state oscillations. In this section, we analyze the stability of such steady states by examining the associated Lyapunov exponents.

The saturated steady-state mode amplitudes in both resonators, $\tilde{a}_{1,2}$, are directly associated with a steady-state oscillation frequency $\omega$ and saturated gain value $g_{\rm s}$. If the system is slightly perturbed, we assume that there is a small deviation from the steady-state regime denoted by $\rho_{1,2,} \propto e^{\lambda t}$ where $\lambda$ is the Lyapunov exponent. Our goal is to determine whether these perturbations vanish over time, thereby assessing the stability of the steady-state regime within a small neighborhood of perturbations. Therefore, the signals are described by 
\begin{gather}
a_{1}(t) = \left(\tilde{a}_1 + \rho_1(t)\right)e^{j\omega t}, \label{eq:a1}\\
a_{2}(t) = \left(\tilde{a}_2 + \rho_2(t)\right)e^{j\omega t}.   
\end{gather}
Also, we assume that the saturable nonlinear gain is given in \eqref{eq:NonLGainModel}. 
For this analysis, we linearize the gain model around the steady-state response using Taylor expansion for the small perturbation $\rho_1$. Neglecting the quadratic terms of $\rho_1$, this procedure leads to

\begin{equation}
\begin{split}
g(|a_1|) & = -\gamma_{\mathrm{i}}+\frac{g_0}{1+c(\tilde{a}_1+\rho_1)(\tilde{a}_1+\rho_1)^{*}} \\
& \approx -\gamma_{\mathrm{i}}+\frac{g_0}{1+ c|\tilde{a}_1|^2 + c\left(\tilde{a}_1\rho_1^*+\tilde{a}_1^*\rho_1\right) } \\
& \approx -\gamma_{\mathrm{i}} + \frac{g_0}{1+c|\tilde{a}_1|^2} \left(1 - c\frac{\tilde{a}_1\rho_1^*+\tilde{a}_1^*\rho_1}{1+c|\tilde{a}_1|^2} \right) \\
& = g(|\tilde{a}_1|) + \left(\left. \frac{dg}{d|a_1|^2} \right|_{\tilde{a}_1} \right) \left(\tilde{a}_1\rho_1^*+\tilde{a}_1^*\rho_1\right).    
\end{split}
\label{eq:linearized_gain}
\end{equation}
Using the signal representation in \eqref{eq:a1}, the differential equation for the first resonator \eqref{eq:HE1} becomes
\begin{equation}
    \frac{d \rho_1}{d t} +j\omega(\tilde{a}_1 + \rho_1)=\left[j\omega_{1}+g(|a_1|)\right]\left(\tilde{a}_1 + \rho_1 \right) 
    -j\kappa \left(\tilde{a}_2 + \rho_2 \right).
\end{equation}
In this equation, we substituting $g$ with the linearized gain from \eqref{eq:linearized_gain} leading to
\begin{equation}
\begin{split}
    \frac{d \rho_1}{d t} = \left[j\left(\omega_{1} - \omega \right) + g(|\tilde{a}_1|) + \left(\left. \frac{dg}{d|a_1|^2} \right|_{\tilde{a}_1} \right)  \left(\tilde{a}_1 \rho^{*}_1+\tilde{a_1^{*}} \rho_1\right) \right] \\
    \left(\tilde{a}_1 + \rho_1 \right) -j\kappa \left(\tilde{a}_2 + \rho_2 \right).
\end{split}
\end{equation}
Using the steady state equation $j\omega \tilde{a}_1 = \left( j\omega_1 + g(|\tilde{a}_1|) \right) \tilde{a}_1  - j\kappa \tilde{a}_2 $, and  and neglecting the quadratic terms of $\rho_1$, we obtain
\begin{equation}
\begin{split}
    \frac{d \rho_1}{d t} = \left[j\left(\omega_{1} - \omega \right) + g(|\tilde{a}_1|) + \left(\left. \frac{dg}{d|a_1|^2} \right|_{\tilde{a}_1} \right) |\tilde{a}_1|^2 \right]\rho_1 \\
    \left(\left. \frac{dg}{d|a_1|^2} \right|_{\tilde{a}_1} \right) \tilde{a}_1^2 \rho_1^*   -j \kappa \rho_2.
\end{split}
\end{equation}

After applying the same procedure to the second resonator, the linearized differential equations for both resonators are given by
\begin{gather}
\frac{d}{d t} \rho_{1} = A \rho_{1} + B \rho_{1}^* + C \rho_{2}, \\
\frac{d}{d t} \rho_{2} = C \rho_{1} + D \rho_{2}, 
\end{gather}
where
\begin{gather}
A=j\left(\omega_{1}-\omega\right) + g(|\tilde{a}_1|) + \left(\left. \frac{dg}{d|a_1|^2} \right|_{\tilde{a}_1} \right) |\tilde{a}_1|^2, \\
B=\left(\left. \frac{dg}{d|a_1|^2} \right|_{\tilde{a}_1} \right) \tilde{a}_1^2, \\
C=-j\kappa, \quad
D=j\left(\omega_2-\omega\right)-\gamma.
\end{gather}

Following Ref. \cite{Zhou2016_PTSymmBreak}, we assume exponential time dependence of the perturbations as,%
\begin{gather}
    \rho_{1} = u_{1} e^{\lambda t}+v_{1}^{*} e^{\lambda^{*} t}, \\
    \rho_{2} = u_{2} e^{\lambda t}+v_{2}^{*} e^{\lambda^{*} t}.
\end{gather}    
The resulting linear eigenvalue problem is written in matrix form as
\begin{equation}
\left[\begin{array}{cccccc}
A & B & C & 0  \\
B^{*} & A^{*} & 0 & C^{*} \\
C & 0 & D & 0 \\
0 & C^{*} & 0 & D^{*} \\
\end{array}\right]\left[\begin{array}{c}
u_{1} \\
v_{1} \\
u_{2} \\
v_{2}
\end{array}\right]=\lambda\left[\begin{array}{c}
u_{1} \\
v_{1} \\
u_{2} \\
v_{2}
\end{array}\right].
\label{eq:Diff_Matrix}
\end{equation}

Each unique steady-state solution of the coupled oscillators will thus have four associated eigenvalues ($\lambda$). Asymptotic stability is guaranteed for a steady-state when all four eigenvalues have negative real part. However, for this system, one Lyapunov exponent equals zero due to an unknown global phase \cite{Assawaworrarit2017_robust,cerjan2013steady}, as the absolute phase of each oscillator in steady-state depends on the initial conditions and the transient, thus the stability is guaranteed when three of the eigenvalues are negative, and one is zero. In order to calculate the four eigenvalues ($\lambda$) for a given set of parameters $\gamma,\kappa,\omega_1,\omega_2, g_0, \gamma_\mathrm{i}$ and $c$ the $\omega, g_{\rm s}$ and $\tilde{a}_1$ values must all be found through using the methods in Sec.~\ref{sec:GeneralNLHamiltonian} and through using the definition of the saturated energy, (\ref{eq:atilde}). One additional caveat is that for a stable state to exist, the $\tilde{a}_1$ must be proper, meaning that $g_0 - \gamma_i \ge g_{\rm s}$.

We show the stability of the previously plotted steady-state solutions from Fig.~\ref{fig:SymmetricSolutions}, Fig.~\ref{fig:ToySol}, and Fig.~\ref{fig:ToySolGs} in Fig.~\ref{fig:StabilityAnalysis}. In the regions of only a single steady state oscillation frequency, contained in parts of all plots, the steady-state solution is stable. In the region of three steady-state solutions, which region is contained in plots (a), (b), and (e), only one or two of the steady-state solutions are stable. The third-order SS-EPD is an interesting point in the parameter space, contained at the center of plots (a) and (d), at which multiple Lyapunov exponents equal zero, indicating neither asymptotic stability nor a diverging unstable solution.

\section{RLC circuit approximations} \label{Appendix:RLCApprox}

\subsection{Expressing RLC oscillator in coupled-mode theory (CMT) terms} \label{Appendix:ExactCMT}

Here we apply CMT to a single RLC oscillator, providing the basics for finding the approximations of the coupled system. The circuit equation for a single parallel RLC oscillator is
\begin{equation}\label{eq:CircEqBasic}
    \frac{d^2 Q}{dt^2} - \frac{G}{C}\frac{dQ}{dt}+\frac{1}{LC}Q =0,
\end{equation}
where $G$ is the resistor's conductance, $L$ is the inductance, $C$ is the capacitance, and $Q$ is the charge accumulated in the capacitor.
The CMT equation for the same circuit is
\begin{equation}\label{eq:CMTBasic}
    \frac{d a}{dt} = (j\omega_0 - \gamma)a.
\end{equation}
First, we rewrite (\ref{eq:CircEqBasic}) in operator form,
\begin{equation}\label{eq:BaseOpperator}
    \left(\frac{d^2 }{dt^2} - \frac{G}{C}\frac{d}{dt}+\frac{1}{LC}\right) Q =0
\end{equation}
that is rewritten as $b_-b_+Q=0$, where 
\begin{equation}\label{eq:LadderEq}
\begin{split}
    b_+ =\frac{d }{dt} - \left(\frac{G}{2C} + j\sqrt{\frac{1}{LC}-\frac{G}{2C}}\right),  \\
    b_- =\frac{d }{dt} - \left(\frac{G}{2C} - j\sqrt{\frac{1}{LC}-\frac{G}{2C}}\right).
\end{split}
\end{equation}
The two equation $b_+ Q=0$ and $b_- Q=0$ lead to positive and negative frequencies, respectively. Equating $b_+ Q=0$ and (\ref{eq:CMTBasic}), finds the exact values of $\gamma = \frac{G}{2C}$ and $\omega_0 = \sqrt{\frac{1}{LC}-\gamma}$ \cite{Haus1984_WavesFields}.

\subsection{Inductively coupled circuits Hamiltonian approximation derivation} \label{Appendix:IndCoupledCirc}

We use the same methodology at the previous section to estimate the relation between the CMT coefficients in (\ref{eq:HE1}) and (\ref{eq:HE2}) and the inductively coupled circuit. The circuit equations governing the inductively coupled circuits in Fig.~\ref{fig:IndCirFig}(a) are
\begin{equation}  \label{eq:C1E1}
\frac{d^2 Q_1}{dt^2} = -\frac{1}{LC_1(1-k^2)}Q_1 + \frac{k}{LC_2(1-k^2)}Q_2 + \frac{G_1}{C_1} \frac {dQ_1}{dt},
\end{equation}
\begin{equation}  \label{eq:C1E2}
\frac{d^2 Q_2}{dt^2} = \frac{k}{LC_1(1-k^2)}Q_1 - \frac{1}{LC_2(1-k^2)}Q_2 - \frac{G_2}{C_2} \frac {dQ_2}{dt},
\end{equation}
where $k=M/L$, and $M$ is the mutual inductance.

It us useful to write the two equations in operator form and use also the positive and negative frequency operators $b_{G\pm} = \left( \frac{d}{dt} - \gamma_1 \mp j\sqrt{\frac{1}{LC_1(1-k^2)} - \gamma_1^2}\right)$, and $b_{L\pm} = \left( \frac{d}{dt} + \gamma_2  \mp j \sqrt{\frac{1}{L C_2(1-k^2)} - \gamma_2^2} \right)$, with $\gamma_i = \frac{G_i}{2 C_i}$. The operator versions of (\ref{eq:C1E1}) and (\ref{eq:C1E2}) become, 
\begin{equation}\label{eq:IndCoupledOpperatorEq}
\begin{split}
    b_{G+}b_{G-}Q_1 - \frac{k\omega_1}{1-k^2}Q_2=0  \\
    b_{L+}b_{L-}Q_2- \frac{k\omega_2}{1-k^2} Q_1   =0.
\end{split}
\end{equation}
These equations cannot be used to directly derive the values for $\omega_1,\omega_2, \gamma,$ and $\kappa$ found in (\ref{eq:HE1}) and (\ref{eq:HE2}) as the negative frequency operator cannot be separated from the positive frequency oscillator. However, when $k\ll1$ this system's positive frequencies approximately behaves as $b_{G+}Q_1=0$ and $b_{L+}Q_2=0$. This is better seen solving for $Q_1$ or $Q_2$ from (\ref{eq:IndCoupledOpperatorEq}). Combining these equation and solving for $Q_1$ this becomes
\begin{equation}\label{eq:CoupledOpperatorSolved}
    \left(b_{L+}b_{L-}b_{G+}b_{G-} - \frac{k^2\omega_1\omega_2}{\left(1-k^2\right)^2}\right)Q_1=0.
\end{equation}
Thus, for small $k$, the approximate CMT parameters $\omega_1,\omega_2, \gamma,$ and $\kappa$ are taken by comparing $b_{G+}Q_1=0$ and $b_{L+}Q_2=0$ with (\ref{eq:HE1}) and (\ref{eq:HE2}). This comparison leads to the approximations recorded in (\ref{eq:InductiveCoupledFirstApprox})-(\ref{eq:InductiveCoupledLastApprox}).

\subsection{Capacitively coupled circuits Hamiltonian approximation derivation} \label{Appendix:CapCoupledCirc}

The approximated values of  $\omega_1,\omega_2, \gamma,$ and $\kappa$ are found through the same steps as in the previous two sections. The circuit equations for the capactively coupled circuits in Fig.~\ref{fig:CapCirFig}(a) are
\begin{equation}  \label{eq:C2E1}
A\frac{d^2 Q_1}{dt^2} = -\frac{B_2}{LC_1}Q_1 - \frac{C_c}{L C_2^2}Q_2 + \frac{G B_2}{C_1} \frac {dQ_1}{dt} - \frac{G C_c}{C_2^2}\frac {dQ_2}{dt},
\end{equation}
\begin{equation}  \label{eq:C2E2}
A \frac{d^2 Q_2}{dt^2} = -\frac{C_c}{LC_1^2}Q_1 - \frac{B_1}{L C_2}Q_2 + \frac{G C_c}{C_1^2} \frac {dQ_1}{dt} - \frac{G B_1}{C_2}\frac {dQ_2}{dt},
\end{equation}
The positive and negative frequency operators are  $b_{G\pm} = \left( \frac{d}{dt} - \frac{GB_2}{2 A C_1} \mp j\sqrt{\frac{B_2}{A L C_1} - \left(\frac{GB_2}{2 A C_1}\right)^2}\right)$, and $b_{L\pm} = \left( \frac{d}{dt} - \frac{GB_1}{2 A C_2} \mp j\sqrt{\frac{B_1}{A L C_2} - \left(\frac{GB_1}{2 A C_2}\right)^2}\right)$. The simplified equation is thus 
\begin{equation}\label{eq:CapCoupledOpperatorEq}
\begin{split}
    b_{G+}b_{G-}Q_1 + \frac{C_c}{C_2}\left(\frac{G}{AC_2}\frac{d}{dt} + \frac{1}{ALC_2} \right) Q_2=0  \\
    b_{L+}b_{L-}Q_2 + \frac{C_c}{C_1}\left(-\frac{G}{AC_1}\frac{d}{dt} + \frac{1}{ALC_1} \right) Q_1 =0.
\end{split}
\end{equation}

Once again, we cannot directly derive the values for $\omega_1,\omega_2, \gamma,$ and $\kappa$ found in (\ref{eq:HE1}) and (\ref{eq:HE2}) from these equations. However, observationally when $C_c/C_1\ll1$ and $C_c/C_2\ll1$, the approximations of the CMT parameters are found by comparing  $b_{G+}Q_1=0$ and $b_{L+}Q_2=0$ with Eqs.~(\ref{eq:HE1}) and (\ref{eq:HE2}). This comparison leads to the approximations recorded in (\ref{eq:CapCoupledFirstApprox})-(\ref{eq:CapCoupledLastApprox}).
Other approximations are also possible.

\bibliography{First_Nonlinear_EPD_Paper}

\end{document}